\documentclass[printer]{aa}
\usepackage{graphicx,natbib}
\usepackage{txfonts}
\begin{document}
 
%
%
\def\valid{}    

\font\caps=cmcsc10                  
\font\dunh=cmdunh10  at 12.0 true pt 
\font\dunhs=cmdunh10 
\font\vbold=cmbx10 scaled \magstep1 
\font\sevenbf=cmbx7
\font\sevenit=cmti7
\font\Kapi=cmr17

\def\MEV{DOME}
\def\RTE{equation of radiative transfer}
\def\etal{{et al}}
\def\HW{H\&W}
\def\OK{O\&K}
\def\ok{O\&K}
\def\RH{R\&H}

\def\ibmrs{\hbox{\tt RS/6000}}
\def\hp{\hbox{\tt HP~9000}}
\def\dec{\hbox{\tt DEC~5000}}
\def\axp{\hbox{\tt AXP}}
\def\ibmmf{\hbox{\tt IBM~3090}}
\def\ibmpc{\hbox{\tt 486DX}}
\def\cray{\hbox{\tt Cray 2}}
\def\ymp{\hbox{\tt YMP}}
\def\nec{\hbox{\tt NEC}}

\def\g{\gamma}
\def\b{\beta}
\def\m{\mu}
\def\e{\epsilon}
\def\n{\nu}
\def\l{\lambda}
\def\L{\Lambda}
\def\K{{\rm K}}
\def\logg{{\log(g)}}
\def\t{\tau}
\def\pder#1#2{{\partial #1 \over \partial #2}}
\def\div#1#2{{#1\over #2}}
\def\rout{\ifmmode{r_{\rm out}}\else\hbox{$r_{\rm out}$}\fi}
\def\tmax{\ifmmode{\tau_{\rm max}}\else\hbox{$\tau_{\rm max}$}\fi}
\def\tstd{\ifmmode{\tau_{\rm std}}\else\hbox{$\tau_{\rm std}$}\fi}
\def\vmax{\ifmmode{v_{\rm max}}\else\hbox{$v_{\rm max}$}\fi}
\def\muE{\ifmmode{\mu_{\rm E}}\else\hbox{$\mu_{\rm E}$}\fi} 
\def\pE{\ifmmode{p_{\rm E}}\else\hbox{$p_{\rm E}$}\fi} 
\def\bmax{\ifmmode{\b_{\rm max}}\else\hbox{$\b_{\rm max}$}\fi}
\def\kms{\hbox{$\,$km$\,$s$^{-1}$}}
\def\ergs{\hbox{$\,$erg$\,$s$^{-1}$}}
\def\kpc{\hbox{$\,$kpc} }
\def\ang{\hbox{\AA}}
\def\Msun{\hbox{$\,$M$_\odot$} }
\def\Lsun{\hbox{$\,$L$_\odot$} }
\def\Teff{\hbox{$\,T_{\rm eff}$} }
\def\alog#1{\times 10^{#1}}
\def\rin{\hbox{$r_{\rm in}$} }
\def\rout{\hbox{$r_{\rm out}$} }

\def\lstar{\ifmmode{\Lambda^*}\else\hbox{$\Lambda^*$}\fi} 
\def\Lstar{\ifmmode{\Lambda^*}\else\hbox{$\Lambda^*$}\fi} 
\def\Rop{\ifmmode{[R_{ij}]}\else\hbox{$[R_{ij}]$}\fi}
\def\Rij{\Rop}
\def\Rji{\ifmmode{[R_{ji}]}\else\hbox{$[R_{ji}]$}\fi}
\def\Rstar{\ifmmode{[R_{ij}^*]}\else\hbox{$[R_{ij}^*]$}\fi}
\def\Rijstar{\Rstar}
\def\Rjistar{\ifmmode{[R_{ji}^*]}\else\hbox{$[R_{ji}^*]$}\fi}
\def\DRji{\ifmmode{[\Delta R_{ji}]}\else\hbox{$[\Delta R_{ji}]$}\fi}
\def\DRij{\ifmmode{[\Delta R_{ij}]}\else\hbox{$[\Delta R_{ij}]$}\fi}

\def\Jb{{\bar J}}
\def\Jbar{{\bar J}}
\def\Jnew{{\bar J_{\rm new}}}
\def\Jold{{\bar J_{\rm old}}}
\def\Jfs{{\bar J_{\rm fs}}}
\def\Snew{{S_{\rm new}}}
\def\Sold{{S_{\rm old}}}
\def\Amat{\mat{A}}             

\def\ns{\ifmmode{N_{\rm s}}          
        \else\hbox{$N_{\rm s}$}\fi}
\def\ion#1{\hbox{ #1}}         

\def\peq{\mathbin{\hbox{$+$}\hbox{$=$}}}

\def\mat#1{{\bf #1}}     
\def\vek#1{{#1}}         

\newcount\eqcount
\eqcount=0
\def
  \nummer{
    \global\advance\eqcount by 1
    (\the\eqcount)
  }

\def
  \numadv{
    \global\advance\eqcount by 1
  }

\def
   \numout#1{
     (\the\eqcount #1)
  }

\def\ivek#1#2{\ifmmode{\vek{I}^{#1}_{#2}}
        \else\hbox{$\vek{I}^{#1}_{#2}$}\fi}

\def\ip#1{\ivek{+}{#1}}      
\def\im#1{\ivek{-}{#1}}      

\def\tmat#1#2{\ifmmode{\mat{t}^{#1}_{#2}}
        \else\hbox{$\mat{t}^{#1}_{#2}$}\fi}
\def\rmat#1#2{\ifmmode{\mat{r}^{#1}_{#2}}
        \else\hbox{$\mat{r}^{#1}_{#2}$}\fi}
\def\bvek#1#2{\ifmmode{\beta^{#1}_{#2}}
        \else\hbox{$\beta^{#1}_{#2}$}\fi}

\def\tpi#1{\tmat{+}{#1}}
\def\tmi#1{\tmat{-}{#1}}
\def\rmi#1{\rmat{-}{#1}}
\def\rpi#1{\rmat{+}{#1}}
\def\bpi#1{\bvek{+}{#1}}
\def\bmi#1{\bvek{-}{#1}}

\def\tp{\tmat{+}{}}          
\def\tm{\tmat{-}{}}          
\def\rmm{\rmat{-}{}}         
\def\rp{\rmat{+}{}}          
\def\bp{\bvek{+}{}}          
\def\bm{\bvek{-}{}}          
\def\tpm{\tmat{\pm}{}}       
\def\rpm{\rmat{\pm}{}}       
\def\bpm{\bvek{\pm}{}}       

\def\lp{\ifmmode{\lambda^+_\tau}           
        \else\hbox{$\lambda^+_\tau$}\fi}
\def\lm{\ifmmode\lambda^-_\tau             
        \else\hbox{$\lambda^-_\tau$}\fi}

%
%
%
%



\def\aasref@jnl#1{{\rm #1}}

\def\aj{\aasref@jnl{AJ}}                   
\def\araa{\aasref@jnl{ARA\&A}}             
\def\apj{\aasref@jnl{ApJ}}                 
\def\apjl{\aasref@jnl{ApJ}}                
\def\apjs{\aasref@jnl{ApJS}}               
\def\ao{\aasref@jnl{Appl.~Opt.}}           
\def\apss{\aasref@jnl{Ap\&SS}}             
\def\aap{\aasref@jnl{A\&A}}                
\def\aapr{\aasref@jnl{A\&A~Rev.}}          
\def\aaps{\aasref@jnl{A\&AS}}              
\def\azh{\aasref@jnl{AZh}}                 
\def\baas{\aasref@jnl{BAAS}}               
\def\jrasc{\aasref@jnl{JRASC}}             
\def\memras{\aasref@jnl{MmRAS}}            
\def\mnras{\aasref@jnl{MNRAS}}             
\def\pra{\aasref@jnl{Phys.~Rev.~A}}        
\def\prb{\aasref@jnl{Phys.~Rev.~B}}        
\def\prc{\aasref@jnl{Phys.~Rev.~C}}        
\def\prd{\aasref@jnl{Phys.~Rev.~D}}        
\def\pre{\aasref@jnl{Phys.~Rev.~E}}        
\def\prl{\aasref@jnl{Phys.~Rev.~Lett.}}    
\def\pasp{\aasref@jnl{PASP}}               
\def\pasj{\aasref@jnl{PASJ}}               
\def\qjras{\aasref@jnl{QJRAS}}             
\def\skytel{\aasref@jnl{S\&T}}             
\def\solphys{\aasref@jnl{Sol.~Phys.}}      
\def\sovast{\aasref@jnl{Soviet~Ast.}}      
\def\ssr{\aasref@jnl{Space~Sci.~Rev.}}     
\def\zap{\aasref@jnl{ZAp}}                 
\def\nat{\aasref@jnl{Nature}}              
\def\iaucirc{\aasref@jnl{IAU~Circ.}}       
\def\aplett{\aasref@jnl{Astrophys.~Lett.}} 
\def\apspr{\aasref@jnl{Astrophys.~Space~Phys.~Res.}}
\def\bain{\aasref@jnl{Bull.~Astron.~Inst.~Netherlands}} 
\def\fcp{\aasref@jnl{Fund.~Cosmic~Phys.}}  
\def\gca{\aasref@jnl{Geochim.~Cosmochim.~Acta}}   
\def\grl{\aasref@jnl{Geophys.~Res.~Lett.}} 
\def\jcp{\aasref@jnl{J.~Chem.~Phys.}}      
\def\jgr{\aasref@jnl{J.~Geophys.~Res.}}    
\def\jqsrt{\aasref@jnl{J.~Quant.~Spec.~Radiat.~Transf.}}
\def\memsai{\aasref@jnl{Mem.~Soc.~Astron.~Italiana}}
\def\nphysa{\aasref@jnl{Nucl.~Phys.~A}}   
\def\physrep{\aasref@jnl{Phys.~Rep.}}   
\def\physscr{\aasref@jnl{Phys.~Scr}}   
\def\planss{\aasref@jnl{Planet.~Space~Sci.}}   
\def\procspie{\aasref@jnl{Proc.~SPIE}}   

\let\astap=\aap
\let\apjlett=\apjl
\let\apjsupp=\apjs
\let\applopt=\ao

\def\phxO{{\tt PHOENIX/1D}}
\def\phxT{{\tt PHOENIX/3D}}
\def\phx{{\tt PHOENIX}}

\baselineskip=12pt

\title{A 3D radiative transfer framework: VI. \phxT\ example applications}

\titlerunning{3D radiative transfer framework VI}
\authorrunning{Hauschildt and Baron}
\author{Peter H. Hauschildt\inst{1} and E.~Baron\inst{1,2,3}}

\institute{
Hamburger Sternwarte, Gojenbergsweg 112, 21029 Hamburg, Germany;
yeti@hs.uni-hamburg.de 
\and
Homer L.~Dodge Dept.~of Physics and Astronomy, University of
Oklahoma, 440 W.  Brooks, Rm 100, Norman, OK 73019 USA;
baron@ou.edu
\and
Computational Research Division, Lawrence Berkeley National Laboratory, MS
50F-1650, 1 Cyclotron Rd, Berkeley, CA 94720-8139 USA
}

\date{Received date \ Accepted date}

\abstract
{}
{We demonstrate the application of our 3D radiative
transfer framework  in the model atmosphere code
\phx\ for a number of spectrum synthesis 
calculations for very different  conditions.
}
{The 3DRT framework discussed in the previous papers of this series was added
to our general-purpose model atmosphere code \phxO\ and an extended 3D version
\phxT\ was created.  The \phxT\ code is parallelized via the  MPI
library using  a hierarchical domain decomposition and displays 
very good strong scaling.
}
{We present the results of several test cases for widely
different atmosphere conditions and compare the 3D calculations
with equivalent 1D models to assess the internal accuracy of the 3D modeling.
In addition, we show the results for a number of parameterized 3D structures.
}
{With presently available computational resources it is possible to solve the
full 3D radiative transfer (including scattering) problem with the same
micro-physics as included in 1D modeling.}


\maketitle

\section{Introduction}

In a series of papers \citet*[][hereafter: Papers I--V]{3drt_paper1,
3drt_paper2, 3drt_paper3, 3drt_paper4, 3drt_paper5}, we have described a
framework for the solution of the radiative transfer equation in 3D systems
(3DRT), including a detailed treatment of scattering in continua and lines with
a non-local operator splitting method. These papers deal solely with the
radiation transport problem and its numerical solution for test cases designed
to stress-test the algorithms and codes. It is important, however, to apply the
radiative transfer codes to `real' problems, e.g., model atmosphere simulations
and to compare the results to 1D equivalents. We have extended our general
purpose model atmosphere code \phx\ to use the 3DRT framework so that the new
version of \phx\ can calculate both 1D (\phxO) and 3D (\phxT) models and
spectra. In this paper we will describe the implementation and the results of
\phx\ calculations comparing the results of 1D and 3D spectrum syntheses for
different model parameters. 

\section{Method}

In the following discussion we use  notation of Papers I -- V.  The
basic framework and the methods used for the formal solution and the solution
of the scattering problem via  non-local operator splitting are discussed in detail in
these papers and will not be repeated here. 

\section{\phxT\ implementation and micro-physics}

We have implemented \phxT\ to use as much as possible of the micro-physics of
\phxO. This applies to the ACES equation of state (Barman, in preparation), to the b-f and
f-f opacities, to dust opacities, and to the line opacities (\phxT\ is
presently restricted to LTE population densities). This includes individual
line profiles (Gauss profiles for weak lines and Voigt profiles for strong
lines depending on user-selectable selection criteria) for atomic and molecular
lines with the same physics that is implemented in \phxO, so that the results
of the opacity calculations are equal for the same physical conditions for
the two modes of \phx. 

The important considerations of \phxT\ implementation are memory and CPU time
consumption. The memory requirements of \phxT\ compared to \phxO\ are mostly due
the the much larger number of voxels in the 3D case (typically $10^6$ voxels)
compared to the 1D case (usually 64--128 layers). As the memory required to
store (and to compute) physical data such as the partial pressures of close to
900 species or the opacities scales linearly with the number of cells (or
layers in 1D), it is obvious that only very small tests can be run without
using domain decomposition methods on large scale parallel supercomputers. The
domain decomposition implementation of \phxT\ distributes the task of solving
(and storing) the equation of state data and the wavelength dependent opacities
to sets of processes each with its private memory. This linearly (with number
of processes used) reduces the amount of memory  and time required for these
tasks. For 1024 processes, this reduces the memory requirements to just a few
MB per process to store the full equation of state results. The 3DRT requires,
in comparison, a total of about 450MB for the same problem (due to the storage
requirements of the non-local \Lstar-operator). Including the storage required
for the computation of the line opacities, this is still just about
0.5GB/process, which is small compared to the typically available 4-16GB/core
(CPU) on modern parallel supercomputers. In order to fully utilize the
available memory per core and to increase flexibility we have implemented a
hierarchical scheme similar to the parallel \phxO\ implementation discussed in
\cite{parapap} and in \cite{parapap2} and to the 3DRT parallelization in Paper II: We use a number of
`clusters' of processes where every cluster works on a different wavelength.
Each cluster internally uses (on its subset of processes) the domain
decomposition discussed above and the 3DRT parallelizations discussed in Paper
I. This scheme can be adjusted to (a) fit the problem in the memory available
for each core and (b) to optimize overall performances (e.g., depending on the
number of solid angle points for the 3DRT solution or the coordinate system
used). In the calculations presented here, we typically use clusters with
256--1024 processes, the number of clusters is limited only by the number
of available CPUs.

\section{Results}

We have calculated a number of test models to compare the 
results of \phxO\ calculations with \phxT\ results. This comparison 
can be used to adjust the parameters of the 3D calculations (number 
of voxels or solid angle points) to give an accuracy that is 
acceptable for a given investment in computer time. The models
that we show here were taken from the latest \phxO\ grid (in preparation)
of model atmospheres.  In all stellar models (1D and 3D) we have used the set of
solar abundances given in \cite*{AGS02}.

\subsection{Stellar Models}

\begin{figure*}
\centering
\resizebox{\hsize}{!}{\includegraphics[angle=00]{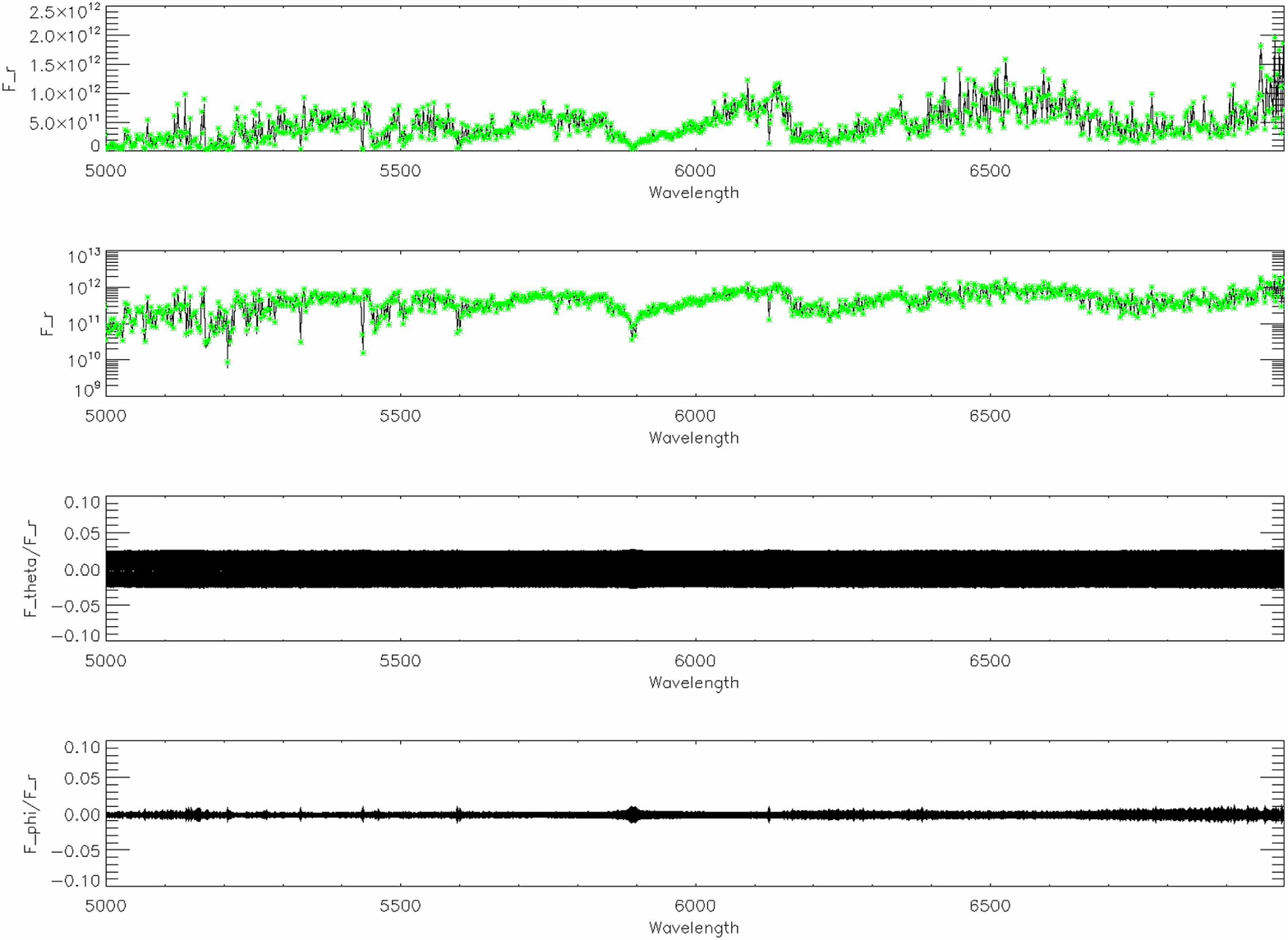}}
\caption{\label{fig:lte30_A}
Comparison between the \phxO\ optical spectrum and the flux vectors across the
outermost voxels for the \phxT\ spectra computed for the M dwarf test model
($\Teff=3000\K$, $\logg=5.0$, `$*$' symbols). 
In the \phxT\ calculations we have used a 3D spherical
coordinate system with $n_r = 129$, $n_{\theta_c}=65$ and $n_{\phi_c}=129$
points for a total of about 1M voxels. The calculations used
$64^2$ solid angle points. 
The top panels show the $F_r$ component of all
outer voxels in linear and logarithmic scales, respectively. The bottom panels
show the corresponding runs of $F_\theta/F_r$ and $F_\phi/F_r$, respectively.
The should be identically zero and the deviations measure the internal
accuracy. See Figs. \ref{fig:lte30_HR} to \ref{fig:lte90_HR} for high-accuracy
solutions for comparison.
The wavelengths are given in {\AA} and the fluxes are in cgs units.
}
\end{figure*}

\begin{figure*}
\centering
\resizebox{\hsize}{!}{\includegraphics[angle=00]{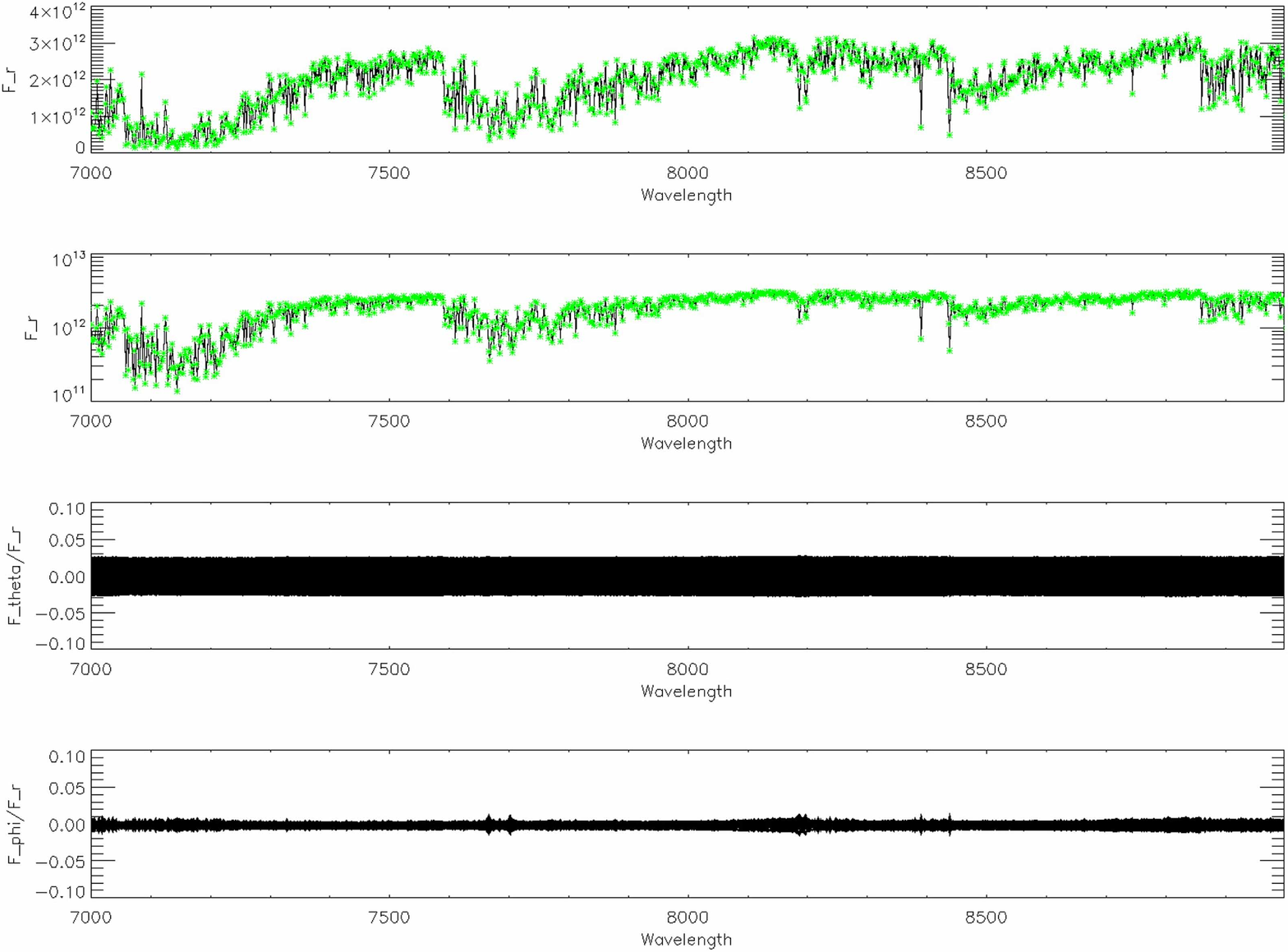}}
\caption{\label{fig:lte30_B}
Comparison between the \phxO\ near infrared spectrum and the flux vectors across the
outermost voxels for the \phxT\ spectra computed for the M dwarf test model
($\Teff=3000\K$, $\logg=5.0$, `$*$' symbols). 
In the \phxT\ calculations we have used a 3D spherical
coordinate system with $n_r = 129$, $n_{\theta_c}=65$ and $n_{\phi_c}=129$
points for a total of about 1M voxels. The calculations used
$64^2$ solid angle points. 
 The top panels show the $F_r$ component of all
outer voxels in linear and logarithmic scales, respectively. The bottom panels
show the corresponding runs of $F_\theta/F_r$ and $F_\phi/F_r$, respectively.
The should be identically zero and the deviations measure the internal
accuracy. See Figs. \ref{fig:lte30_HR} and \ref{fig:lte90_HR} for high-accuracy
solutions for comparison.
The wavelengths are given in {\AA} and the fluxes are in cgs units.
}
\end{figure*}

\begin{figure*}
\centering
\resizebox{\hsize}{!}{\includegraphics[angle=00]{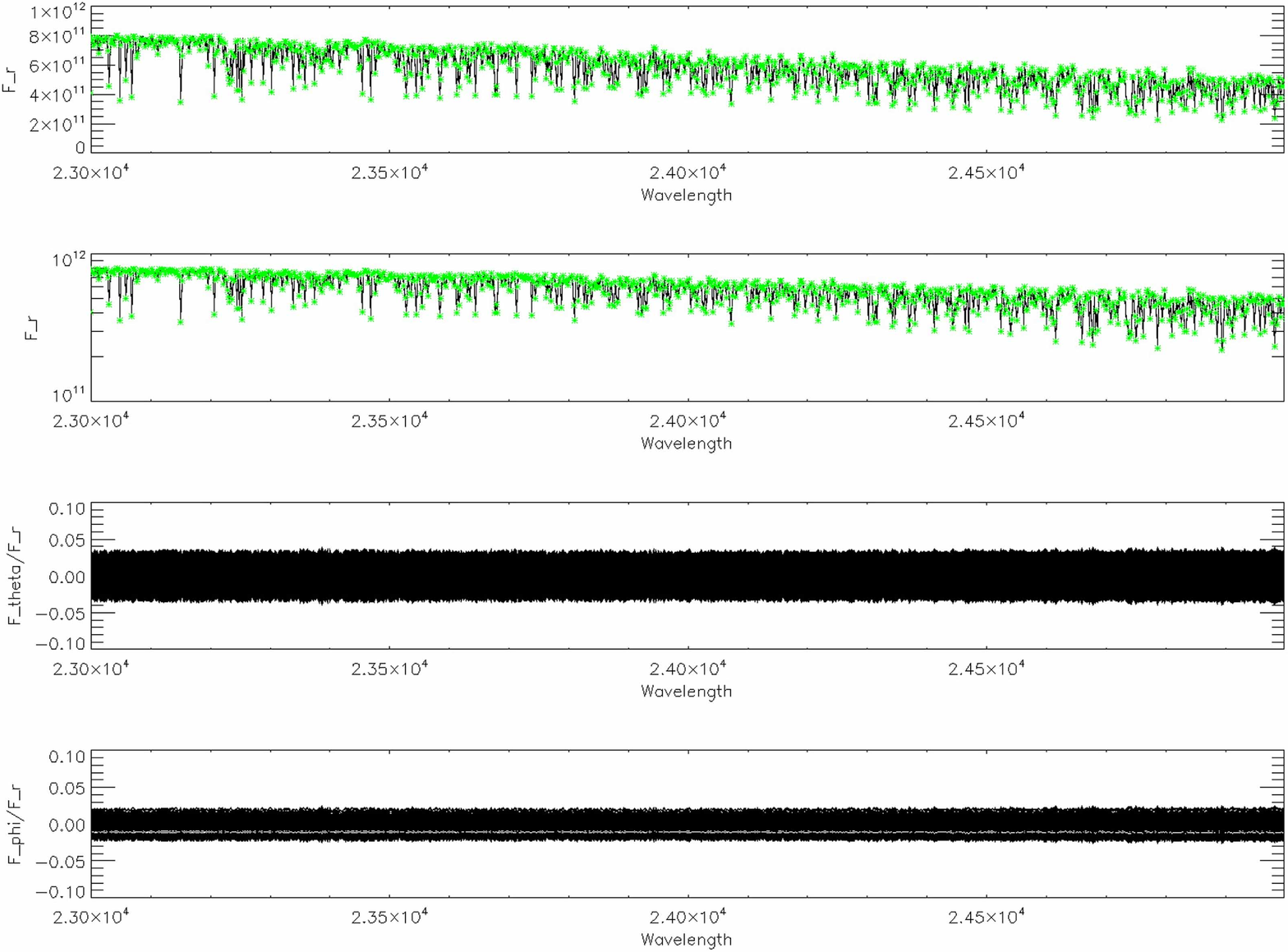}}
\caption{\label{fig:lte30_J} 
Comparison between the \phxO\ infrared spectrum and the flux vectors across the
outermost voxels for the \phxT\ spectra computed for the M dwarf test model
($\Teff=3000\K$, $\logg=5.0$, `$*$' symbols).  
In the \phxT\ calculations we have used a 3D spherical
coordinate system with $n_r = 129$, $n_{\theta_c}=65$ and $n_{\phi_c}=129$
points for a total of about 1M voxels. The calculations used
$64^2$ solid angle points. 
The top panels show the $F_r$ component of all
outer voxels in linear and logarithmic scales, respectively. The bottom panels
show the corresponding runs of $F_\theta/F_r$ and $F_\phi/F_r$, respectively.
The should be identically zero and the deviations measure the internal
accuracy. See Figs. \ref{fig:lte30_HR} and \ref{fig:lte90_HR} for high-accuracy
solutions for comparison.
The wavelengths are given in {\AA} and the fluxes are in cgs units.
}
\end{figure*}

\begin{figure*}
\centering
\resizebox{\hsize}{!}{\includegraphics[angle=00]{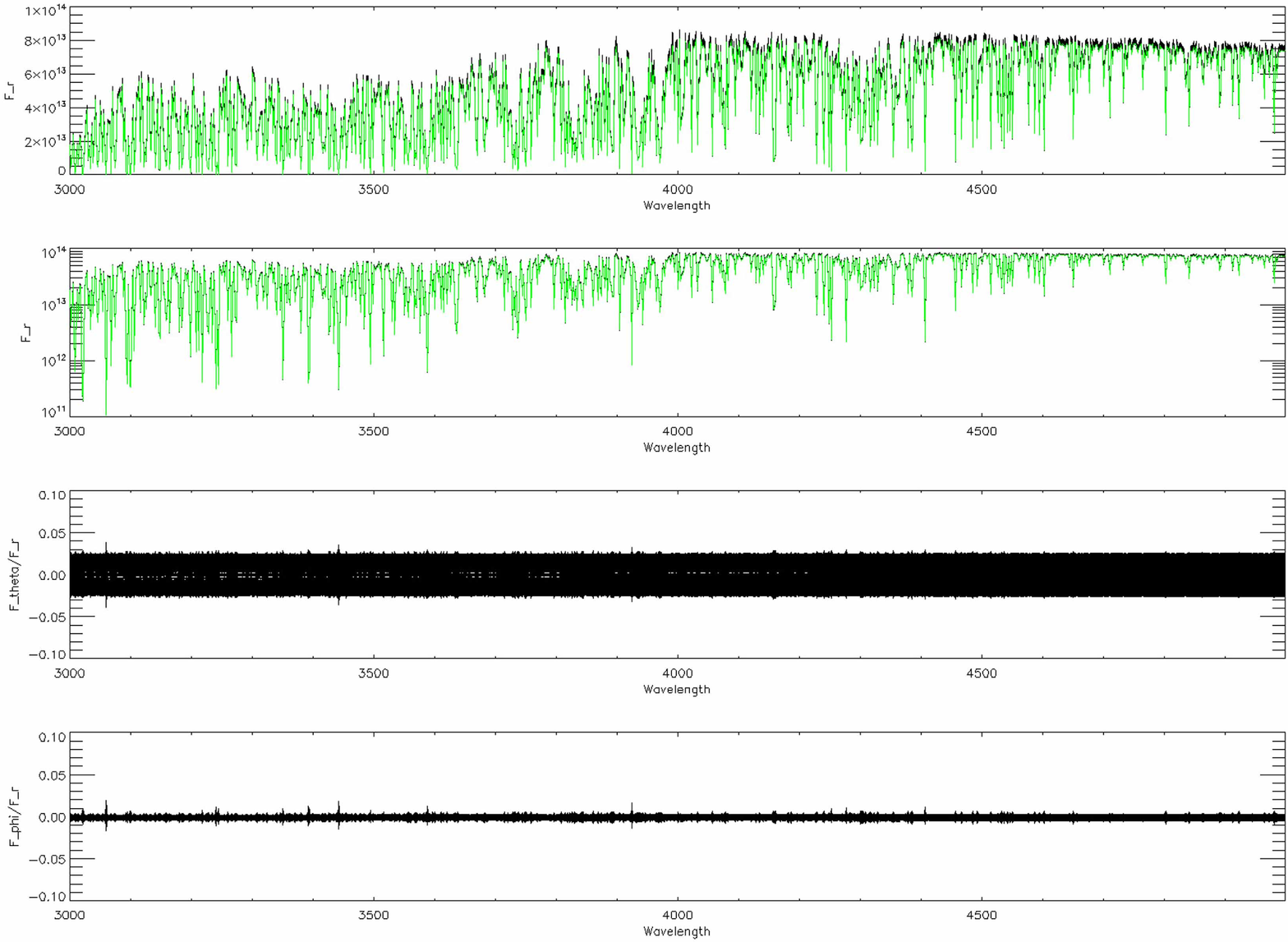}}
\caption{\label{fig:lte57_B}
Comparison between the \phxO\ near UV spectrum and the flux vectors across the
outermost voxels for the \phxT\ spectra computed for the G2V dwarf test model
($\Teff=5700\K$, $\logg=4.5$, `$*$' symbols).  
In the \phxT\ calculations we have used a 3D spherical
coordinate system with $n_r = 129$, $n_{\theta_c}=65$ and $n_{\phi_c}=129$
points for a total of about 1M voxels. The calculations used
$64^2$ solid angle points. 
The top panels show the $F_r$ component of all
outer voxels in linear and logarithmic scales, respectively. The bottom panels
show the corresponding runs of $F_\theta/F_r$ and $F_\phi/F_r$, respectively.
The should be identically zero and the deviations measure the internal
accuracy. See Figs. \ref{fig:lte30_HR} and \ref{fig:lte90_HR} for high-accuracy
solutions for comparison.
The wavelengths are given in {\AA} and the fluxes are in cgs units.
}
\end{figure*}

\begin{figure*}
\centering
\resizebox{\hsize}{!}{\includegraphics[angle=00]{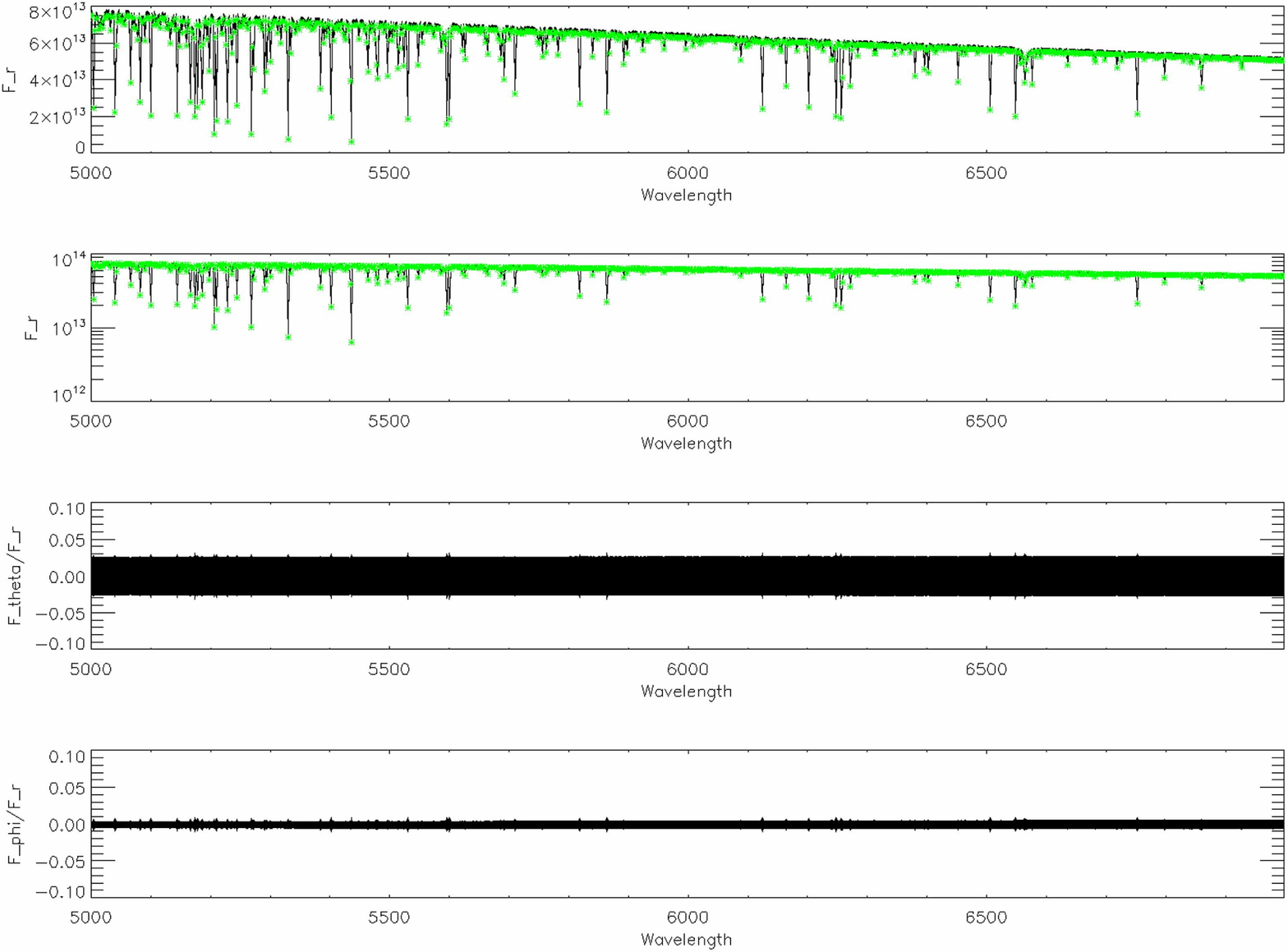}}
\caption{\label{fig:lte57_C}
Comparison between the \phxO\ optical spectrum and the flux vectors across the
outermost voxels for the \phxT\ spectra computed for the G2V dwarf test model
($\Teff=5700\K$, $\logg=4.5$, `$*$' symbols).  
In the \phxT\ calculations we have used a 3D spherical
coordinate system with $n_r = 129$, $n_{\theta_c}=65$ and $n_{\phi_c}=129$
points for a total of about 1M voxels. The calculations used
$64^2$ solid angle points. 
The top panels show the $F_r$ component of all
outer voxels in linear and logarithmic scales, respectively. The bottom panels
show the corresponding runs of $F_\theta/F_r$ and $F_\phi/F_r$, respectively.
The should be identically zero and the deviations measure the internal
accuracy. See Figs. \ref{fig:lte30_HR} and \ref{fig:lte90_HR} for high-accuracy
solutions for comparison.
The wavelengths are given in {\AA} and the fluxes are in cgs units.
}
\end{figure*}

\begin{figure*}
\centering
\resizebox{\hsize}{!}{\includegraphics[angle=00]{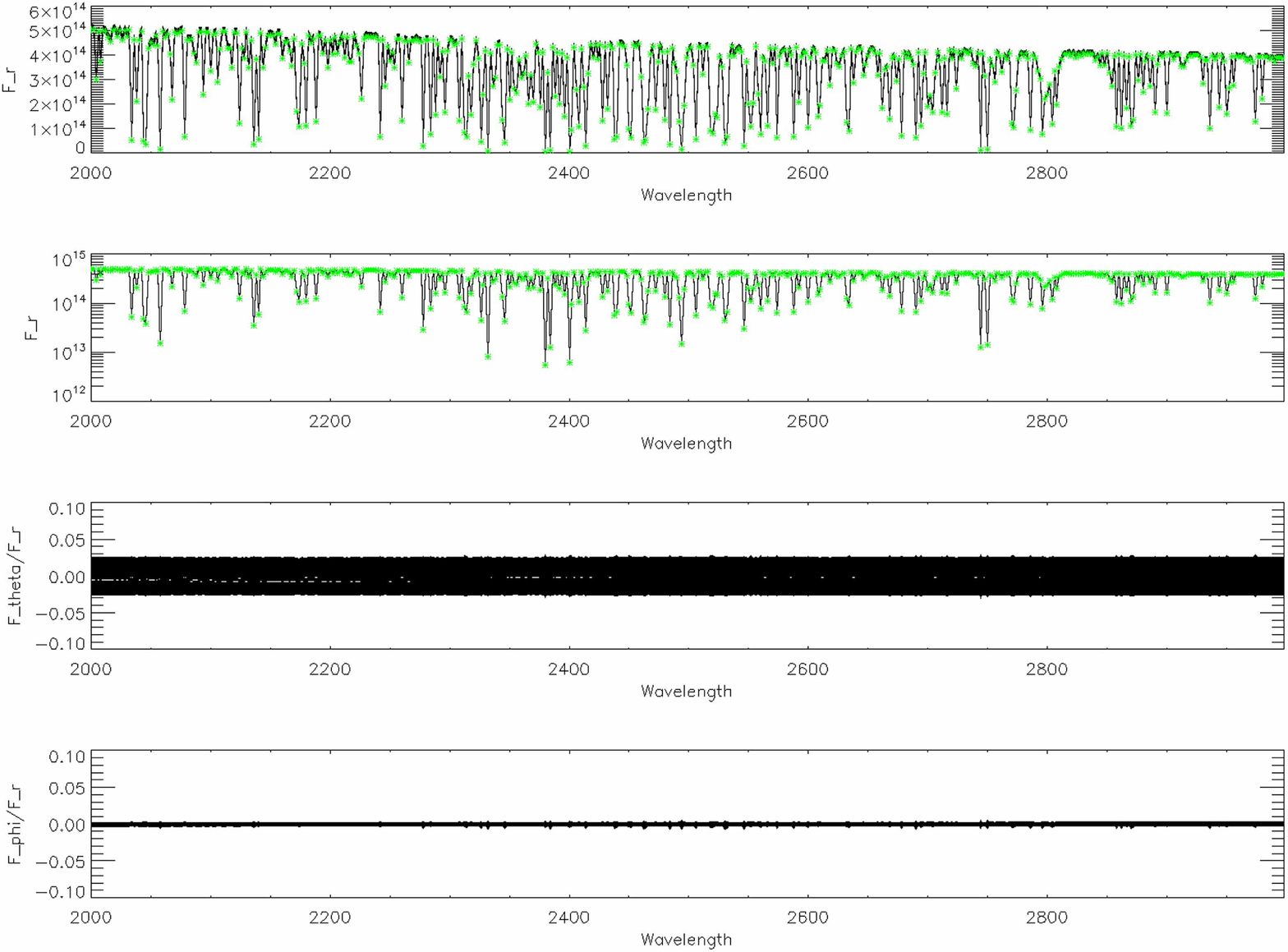}}
\caption{\label{fig:lte90_C} 
Comparison between the \phxO\ optical spectrum and the flux vectors across the
outermost voxels for the \phxT\ spectra computed for the A dwarf test model
($\Teff=9000\K$, $\logg=4.5$, `$*$' symbols).  
In the \phxT\ calculations we have used a 3D spherical
coordinate system with $n_r = 129$, $n_{\theta_c}=65$ and $n_{\phi_c}=129$
points for a total of about 1M voxels. The calculations used
$64^2$ solid angle points. 
The top panels show the $F_r$ component of all
outer voxels in linear and logarithmic scales, respectively. The bottom panels
show the corresponding runs of $F_\theta/F_r$ and $F_\phi/F_r$, respectively.
The should be identically zero and the deviations measure the internal
accuracy. See Figs. \ref{fig:lte30_HR} and \ref{fig:lte90_HR} for high-accuracy
solutions for comparison.
The wavelengths are given in {\AA} and the fluxes are in cgs units.
}
\end{figure*}

\begin{figure*}
\centering
\resizebox{\hsize}{!}{\includegraphics[angle=00]{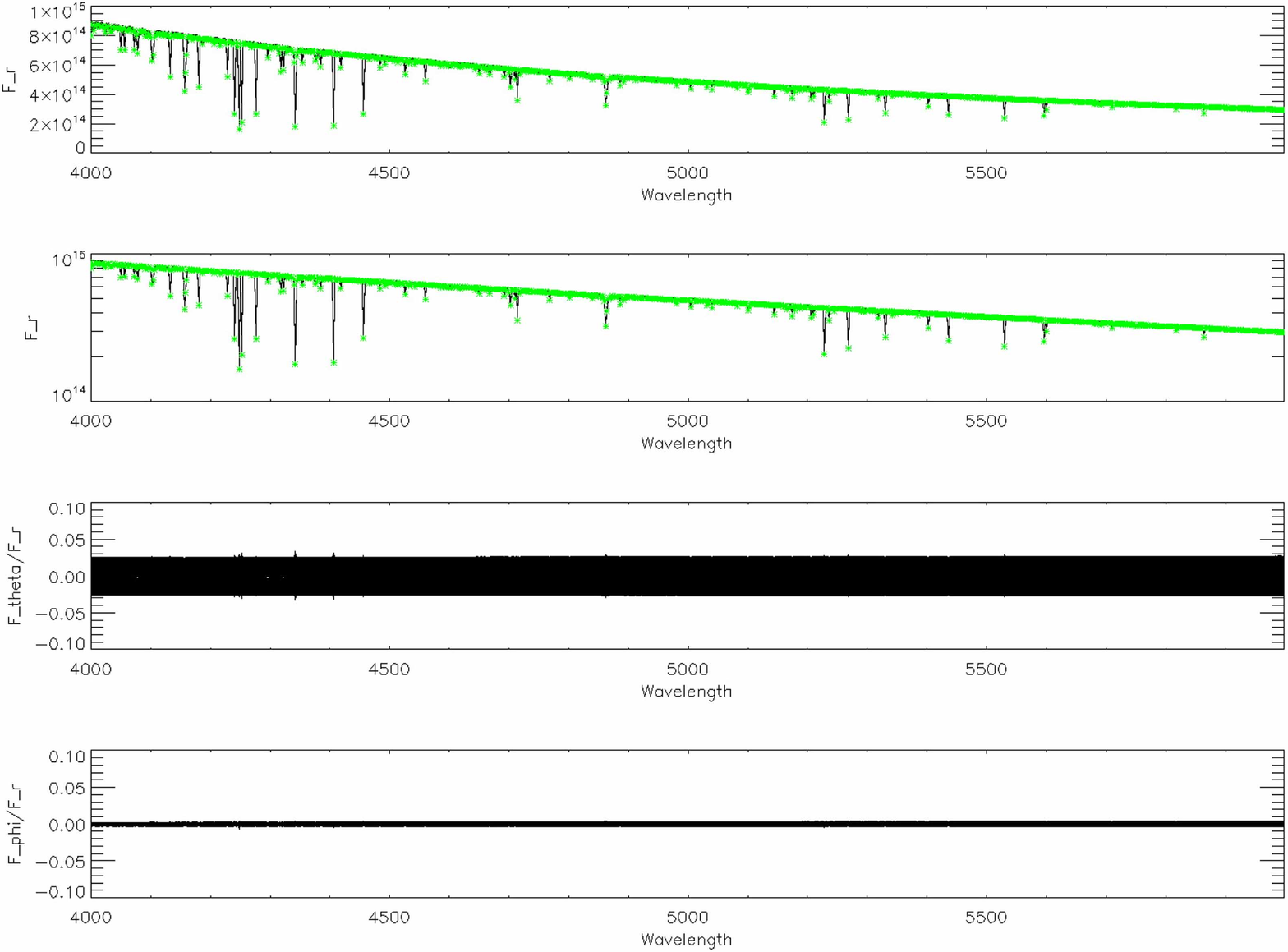}}
\caption{\label{fig:lte90_E}
Comparison between the \phxO\ UV spectrum and the flux vectors across the
outermost voxels for the \phxT\ spectra computed for the A dwarf test model
($\Teff=9000\K$, $\logg=4.5$, `$*$' symbols).  
In the \phxT\ calculations we have used a 3D spherical
coordinate system with $n_r = 129$, $n_{\theta_c}=65$ and $n_{\phi_c}=129$
points for a total of about 1M voxels. The calculations used
$64^2$ solid angle points. 
The top panels show the $F_r$ component of all
outer voxels in linear and logarithmic scales, respectively. The bottom panels
show the corresponding runs of $F_\theta/F_r$ and $F_\phi/F_r$, respectively.
The should be identically zero and the deviations measure the internal
accuracy. See Figs. \ref{fig:lte30_HR} and \ref{fig:lte90_HR} for high-accuracy
solutions for comparison.
The wavelengths are given in {\AA} and the fluxes are in cgs units.
}
\end{figure*}

\begin{figure*}
\centering
\resizebox{\hsize}{!}{\includegraphics[angle=00]{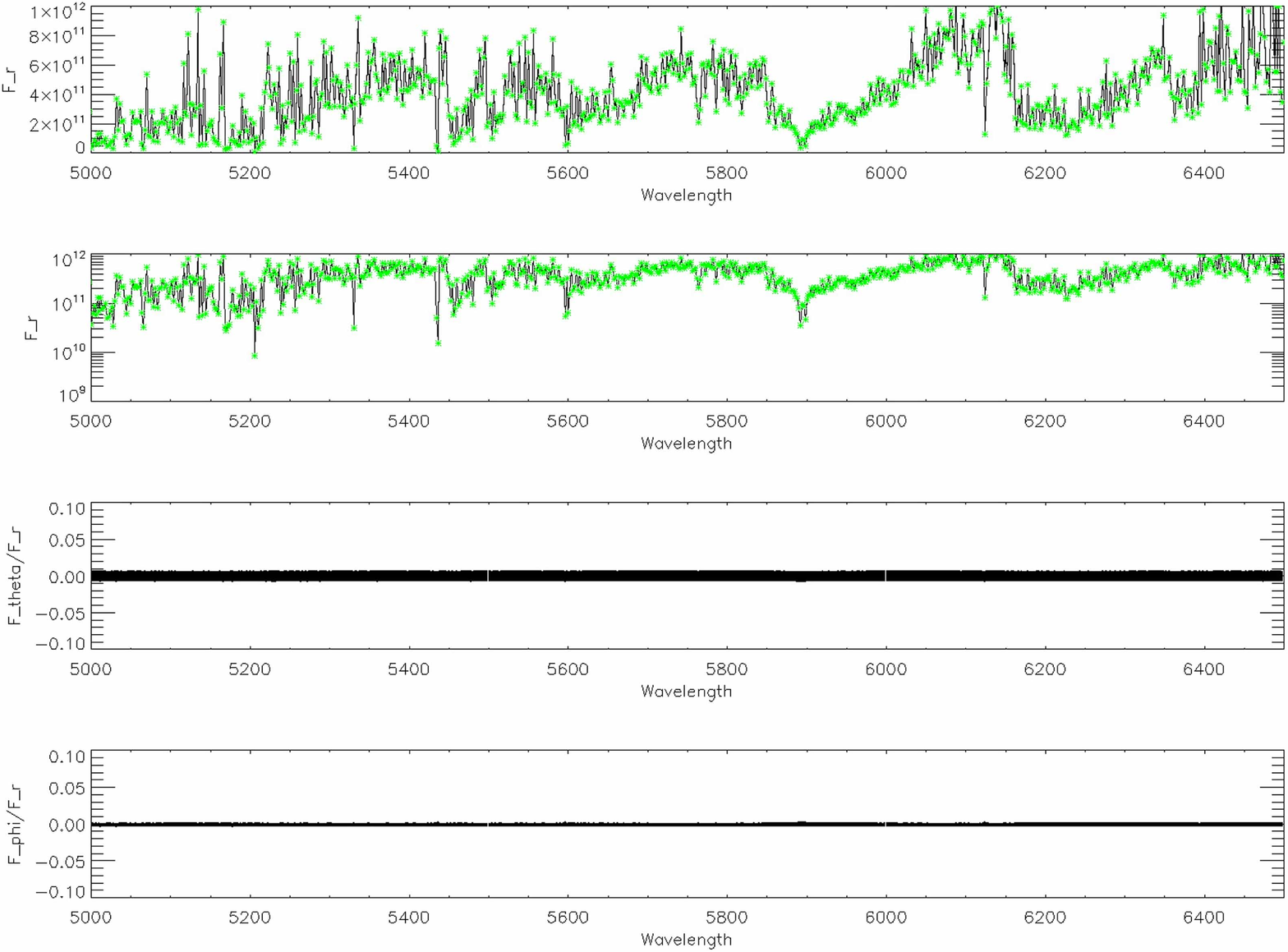}}
\caption{\label{fig:lte30_HR}
Comparison between the \phxO\ optical spectrum and the flux vectors across the
outermost voxels for the higher resolution \phxT\ spectra computed for the M dwarf test model
($\Teff=3000\K$, $\logg=5.0$, `$*$' symbols). 
In the \phxT\ calculations we have used a 3D spherical
coordinate system with $n_r = 129$, $n_{\theta_c}=65$ and $n_{\phi_c}=129$
points for a total of about 1M voxels. The calculations used
$256^2$ solid angle points. 
 The top panels show the $F_r$ component of all
outer voxels in linear and logarithmic scales, respectively. The bottom panels
show the corresponding runs of $F_\theta/F_r$ and $F_\phi/F_r$, respectively.
The should be identically zero and the deviations measure the internal
accuracy.
The wavelengths are given in {\AA} and the fluxes are in cgs units.
}
\end{figure*}

\begin{figure*}
\centering
\resizebox{\hsize}{!}{\includegraphics[angle=00]{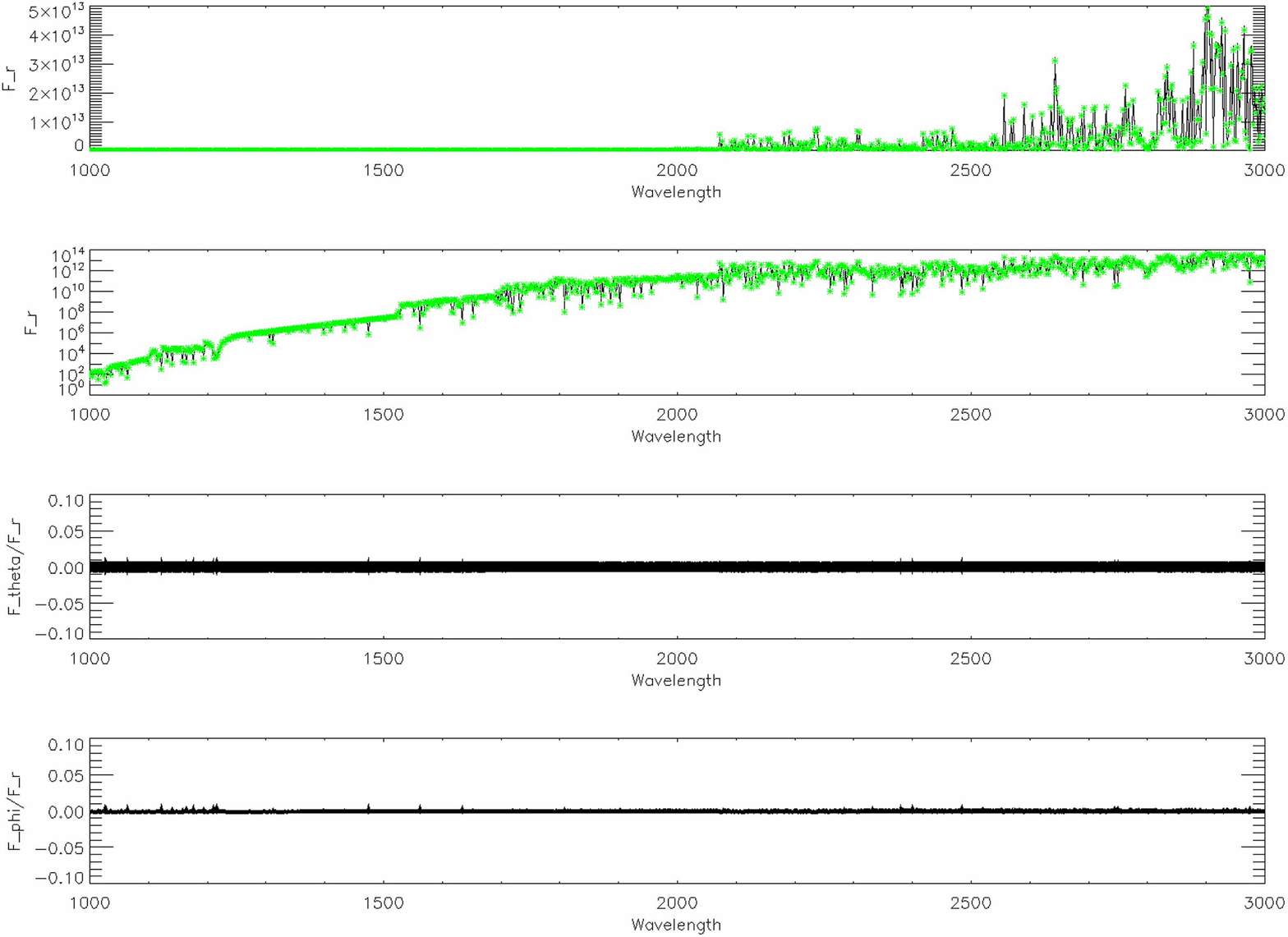}}
\caption{\label{fig:lte57_HR}
Comparison between the \phxO\ UV spectrum and the flux vectors across the
outermost voxels for the higher resolution \phxT\ spectra computed for the G2V dwarf test model
($\Teff=5700\K$, $\logg=4.5$, `$*$' symbols).  
In the \phxT\ calculations we have used a 3D spherical
coordinate system with $n_r = 129$, $n_{\theta_c}=65$ and $n_{\phi_c}=129$
points for a total of about 1M voxels. The calculations used
$256^2$ solid angle points. 
The top panels show the $F_r$ component of all
outer voxels in linear and logarithmic scales, respectively. The bottom panels
show the corresponding runs of $F_\theta/F_r$ and $F_\phi/F_r$, respectively.
The should be identically zero and the deviations measure the internal
accuracy.
The wavelengths are given in {\AA} and the fluxes are in cgs units.
}
\end{figure*}

\begin{figure*}
\centering
\resizebox{\hsize}{!}{\includegraphics[angle=00]{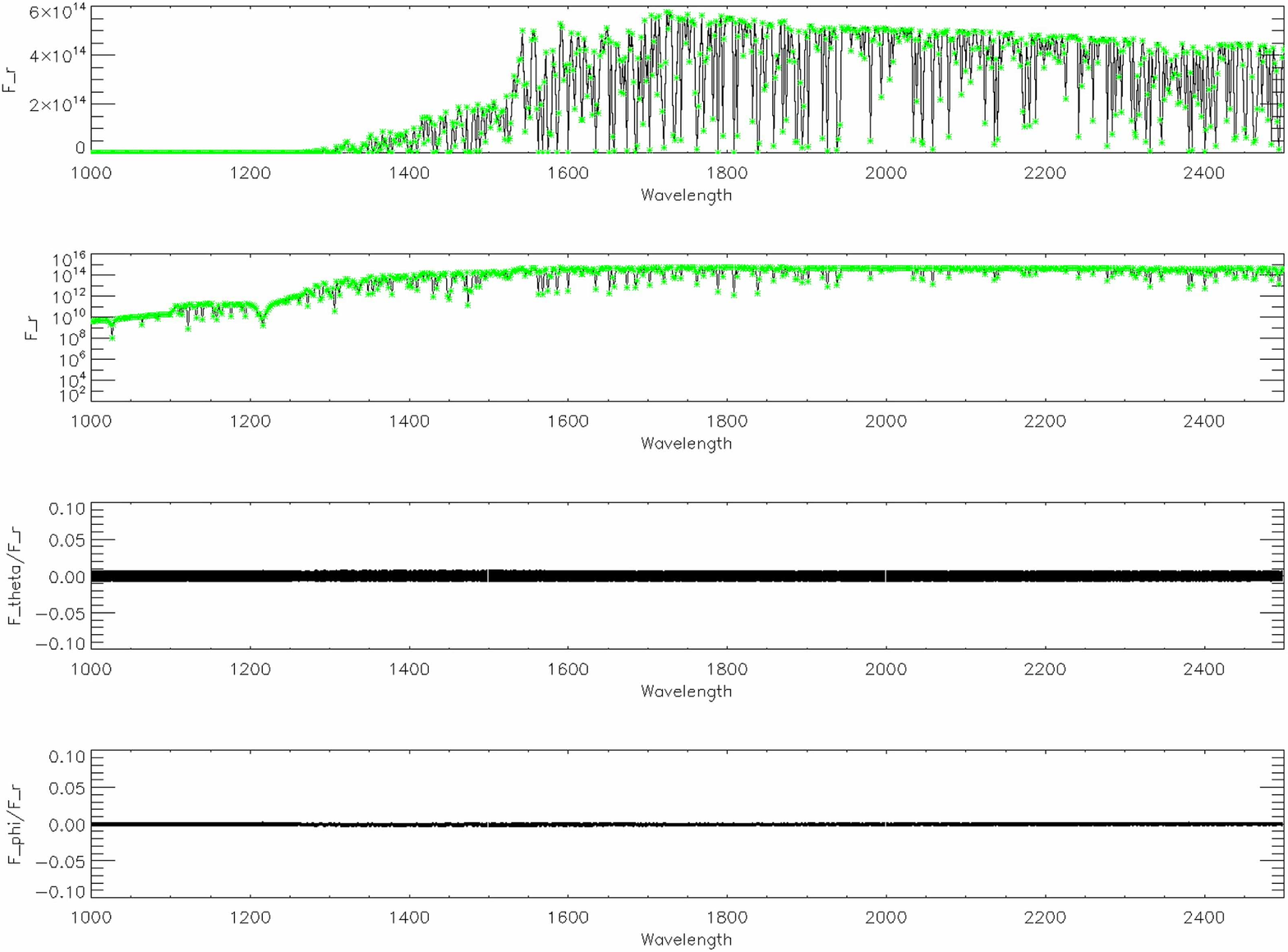}}
\caption{\label{fig:lte90_HR}
Comparison between the optical \phxO\ spectrum and the flux vectors across the
outermost voxels for the higher resolution \phxT\ spectra computed for the A dwarf test model
($\Teff=5700\K$, $\logg=4.5$, `$*$' symbols).  
In the \phxT\ calculations we have used a 3D spherical
coordinate system with $n_r = 129$, $n_{\theta_c}=65$ and $n_{\phi_c}=129$
points for a total of about 1M voxels. The calculations used
$256^2$ solid angle points. 
The top panels show the $F_r$ component of all
outer voxels in linear and logarithmic scales, respectively. The bottom panels
show the corresponding runs of $F_\theta/F_r$ and $F_\phi/F_r$, respectively.
The should be identically zero and the deviations measure the internal
accuracy.
The wavelengths are given in {\AA} and the fluxes are in cgs units.
}
\end{figure*}

We have computed synthetic spectra for stellar model atmospheres
with the parameters $\Teff=3000\K,\ \logg=5.0$ (M dwarf),
$\Teff=5700\K,\ \logg=4.5$ (solar type star) and 
$\Teff=9000\K,\ \logg=4.5$ (A star). The \phxO\ models 
were computed with the latest setup in the input physics,
including the ACES equation of state and  the latest 
version of the atomic and molecular line databases.
The model structures were then used as inputs to 
\phxT\ to calculate synthetic spectra with the same
sampling rates as the spectra from the \phxO\ calculations. 
In the \phxT\ calculations we have used a 3D spherical
coordinate system with $n_r = 129$, $n_{\theta_c}=65$ and $n_{\phi_c}=129$
points for a total of about 1M voxels. The calculations used
(if not specified otherwise) $64^2$ solid angle points. 
For each object we calculated synthetic spectra with \phxO\ and 
\phxT\ and compare the fluxes of the 1D spectra to the flux 
vectors of the 3D results. As in paper IV we can use the 
$(\theta_c,\phi_c)$ components of the 3D flux vector in 
3D spherical coordinates to estimate the internal accuracy of
the solution (as the $F_{\theta_c}$ and $F_{\phi_c}$ components
are zero for spherically symmetric configurations). 
Figures \ref{fig:lte30_A} to \ref{fig:lte90_E} show selected
results for the different models. In these cases, the 
error due to the number of solid angle points is 
about $3$\% and in all tests run the differences
between the \phxO\ fluxes and the $F_r$ component of
the \phxT\ calculation is of the same order. The differences
between the 1D and 3D calculations are within the accuracy
set by the number of solid angles in the 3D model. In order to 
verify that the errors get smaller with larger number of
solid angles (as shown in Paper IV for simple test
cases), we have run test models with $256^2$ angles. Three example plots are 
shown  in Figs.~\ref{fig:lte30_HR} to \ref{fig:lte90_HR} The results show clearly that the higher
solid angle resolution reduces the errors in $F_{\theta_c}$ and $F_{\phi_c}$
considerably and also improves the comparison for $F_r$ to the 1D result,
as the higher internal accuracy due to more solid angle points
also increases the internal accuracy of $F_r$. This also shows
that in 3D radiative transfer calculations the spatial resolution
is not the only factor governing the quality of the solution, the 
solid angle resolution may in fact be more important, depending
on the coordinate system used and the details of the problem
that isve calculated a number of test models to compare the
results of \phxO\ calculations with \phxT\ results. This comparison
can be used to adjust the parameters of the 3D calculations (number
of voxels or solid angle points) to give an accuracy that is
acceptable for a given investment in computer time. The models
that we show here were taken from the latest \phxO\ grid (in preparation)
of model atmospheres.  In all stellar models (1D and 3D) we have used the set of
solar abundances given in \cite*{AGS02}.
We have calculated a number of test models to compare the
results of \phxO\ calculations with \phxT\ results. This comparison
can be used to adjust the parameters of the 3D calculations (number
of voxels or solid angle points) to give an accuracy that is
acceptable for a given investment in computer time. The models
that we show here were taken from the latest \phxO\ grid (in preparation)
of model atmospheres.  In all stellar models (1D and 3D) we have used the set of
solar abundances given in \cite*{AGS02}. being considered.

\subsection{Scaling}

In order to investigate the strong scaling properties of \phxT\ we have constructed a
small test case for a M~dwarf model with 1000 wavelength points in a 3D spherical coordinate system with $n_r = 129$, $n_{\theta_c}=65$ and
$n_{\phi_c}=129$ points for a total of about 1M voxels and $64^2$ solid angle
points and ran the
calculations with different configurations of the domain decomposition and
different total numbers of processes. The total workload remains constant 
in these calculations, so this is a strong scaling test where the workload
per CPU drops as the number of processes increase (in contrast to a weak scaling
test where the workload per process remains constant). The results are given in Table
\ref{tab:1}. In this table, `n(MPI)' is the total number of MPI processes
used, `cluster size' is the number of processes that collaboratively work on a
single wavelength (spatial domain decomposition) and `n(cluster)' is the number
of such clusters, each working on a different wavelength (energy domain
decomposition). The product `cluster size' $\times$ `n(cluster)' is always equal
to `n(MPI)'.  The column `Comm' gives the time spent in MPI communication
to collect the opacities from the different processes before the 3DRT
calculation starts. The communication requirements of the 3DRT calculations are
included in the 3DRT column.
The columns `line opacity' give the time in
seconds and scaling efficiency for all line opacity calculations, respectively,
The columns `total' give the total time and
scaling efficiency, respectively, of the overall time spent in the computation
of the 3D spectrum, this time does not include (small) contributions from the
EOS solution and the line selection procedures.  In the largest cluster size of
512 processes each process only works on 8 solid angles, whereas in the
smallest cluster size (128) each process works on 32 solid angles.  The work
per solid angle is not perfectly constant and the amount of communication
increases linearly as more processes collaborate, therefore, the scaling
efficiency drops if more than about 512 processes are used for this problem
size (i.e., number of solid angles). The scaling efficiency for the overall
problem is quite good, the optimal value is about 98\%. The drop-off for
cluster sizes of 512 (and more) is due to (a) the relatively small number of solid angles
leading to very little work for each 3DRT process and relatively
more internal communication time in the 3DRT and (b) the small effect
of the communication related to the spatial domain decomposition.
We could not test setups with more than (the maximum available) 2048 processes;
however, the test case should scale to 256k processes (number of wavelength points
times cluster size), although for such a setup the overheads for, e.g., 
the solution of the equation of state and the line selection would be 
very noticeable.
\subsection{3D hydro model of solar convection}

For a test with a computed 3D structure, we use the same example snapshot
structure from H-G.\ Ludwig \cite[]{2007A&A...473L...9C,2004A&A...414.1121W} of
a radiation-hydrodynamical simulation of convection in the solar atmosphere as
in Paper III. The radiation transport calculations were performed with a total
of $141\times 141 \times 151$ Cartesian grid points in $x$, $y$, and $z$,
respectively, for a total of $3\,002\,031$ voxels, the periodic boundary
conditions are set in the (horizontal) $x,y$ plane. The 3D radiative transport
equation is solved for $n_{\theta}=64$ and $n_{\phi}=64$ solid angle points, so
that a total of about $12\alog{9}$ intensities are calculated for each 3DRT
iteration and wavelength point. For the tests described here, we are only using
the temperature--pressure structure of the hydro model and ignore the velocity
field. 

\begin{figure*}
\centering
\resizebox{\hsize}{!}{\includegraphics[angle=00]{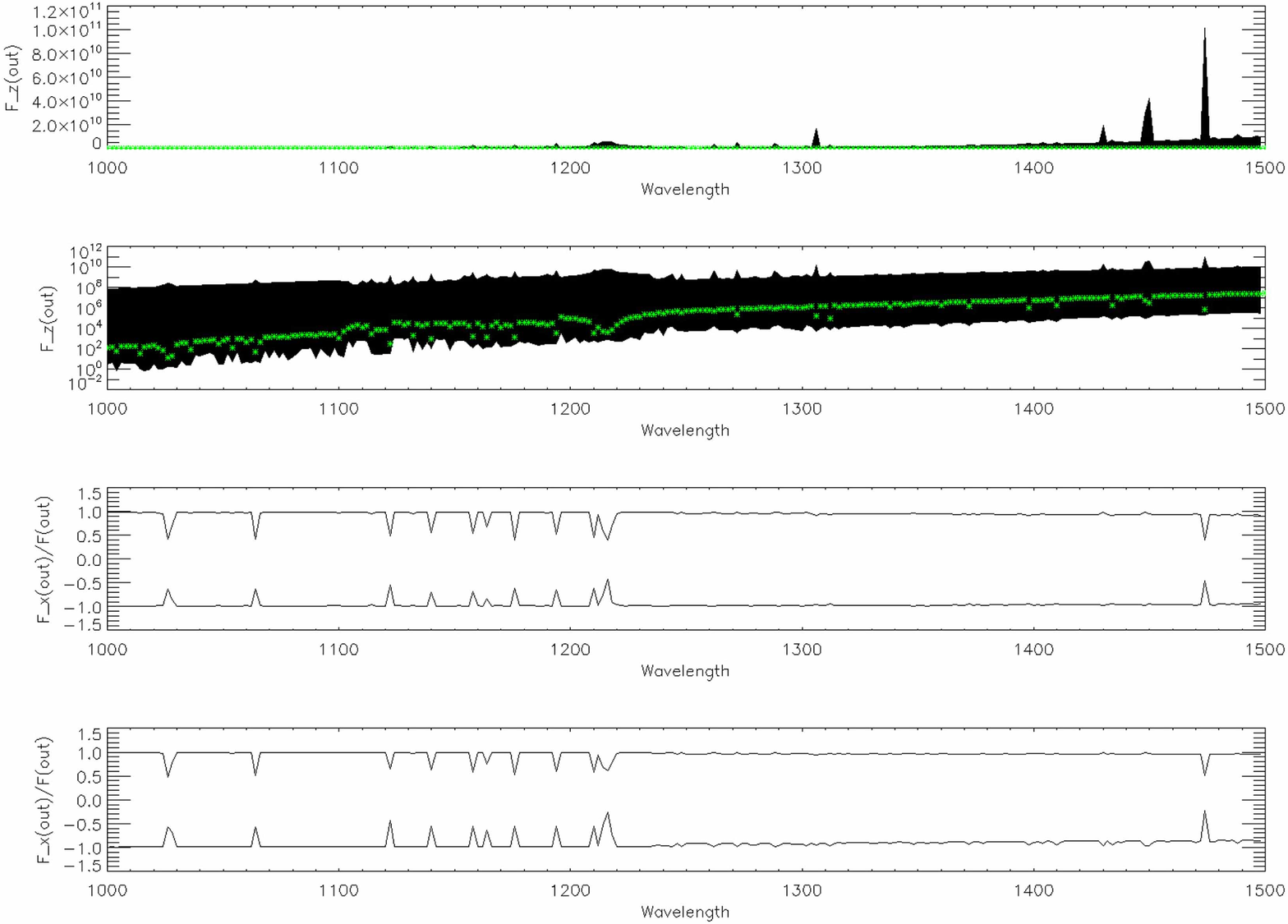}}
\caption{\label{fig:HGL1000}
Comparison between the flux vectors across the outermost voxels for the \phxT\
UV spectra computed for the 3D hydro structure and the \phxO\ spectrum for the G2V
dwarf test model ($\Teff=5700\K$, $\logg=4.5$).  In the \phxT\ calculations we
have used a 3D coordinate system with a total of $141\times 141 \times 151$
Cartesian grid points in $x$, $y$, and $z$, respectively, the periodic boundary
conditions are set in the (horizontal) $x,y$ plane. The 3D radiative transport
equation is solved for $n_{\theta}=64$ and $n_{\phi}=64$ solid angle points.
The top panels show the $F_z$ component of all outer voxels in linear and
logarithmic scales, respectively, compared to the results of the 1D comparison
model. 
{The bottom panels show the corresponding maxima and minima  of $F_x/|\vec F|$ and
$F_y/|\vec F|$, respectively, over all surface voxels for each wavelength.
These panels show that in the 3D structure even at the surface a substantial
horizontal energy flow takes place, see also figures \ref{fig:HGL1090_flow} 
-- \ref{fig:HGL6998_flow}. See text for details.}
The wavelengths are given in {\AA} and the fluxes are in cgs units.
}
\end{figure*}

\begin{figure*}
\centering
\resizebox{\hsize}{!}{\includegraphics[angle=00]{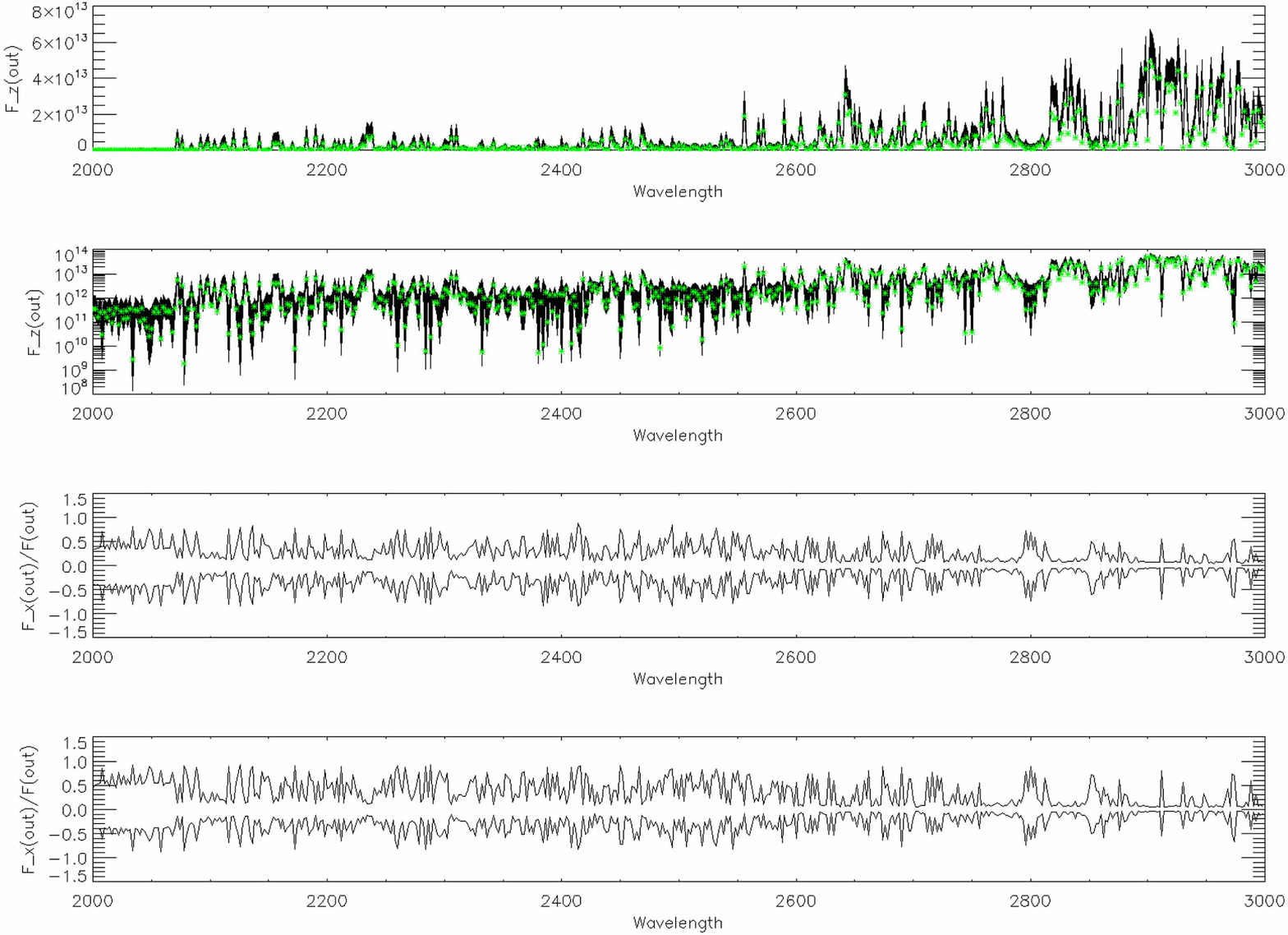}}
\caption{\label{fig:HGL2000}
Comparison between the flux vectors across the outermost voxels for the \phxT\
near UV spectra computed for the 3D hydro structure and the \phxO\ spectrum for the G2V
dwarf test model ($\Teff=5700\K$, $\logg=4.5$).  In the \phxT\ calculations we
have used a 3D coordinate system with a total of $141\times 141 \times 151$
Cartesian grid points in $x$, $y$, and $z$, respectively, the periodic boundary
conditions are set in the (horizontal) $x,y$ plane. The 3D radiative transport
equation is solved for $n_{\theta}=64$ and $n_{\phi}=64$ solid angle points.
The top panels show the $F_z$ component of all outer voxels in linear and
logarithmic scales, respectively, compared to the results of the 1D comparison
model.
{The bottom panels show the corresponding maxima and minima  of $F_x/|\vec F|$ and
$F_y/|\vec F|$, respectively, over all surface voxels for each wavelength.
Note the difference between this result and that shown in the bottom two panels
of Fig.~\ref{fig:HGL1000}. See text for details.
}
The wavelengths are given in {\AA} and the fluxes are in cgs units.
}
\end{figure*}

\begin{figure*}
\centering
\resizebox{\hsize}{!}{\includegraphics[angle=00]{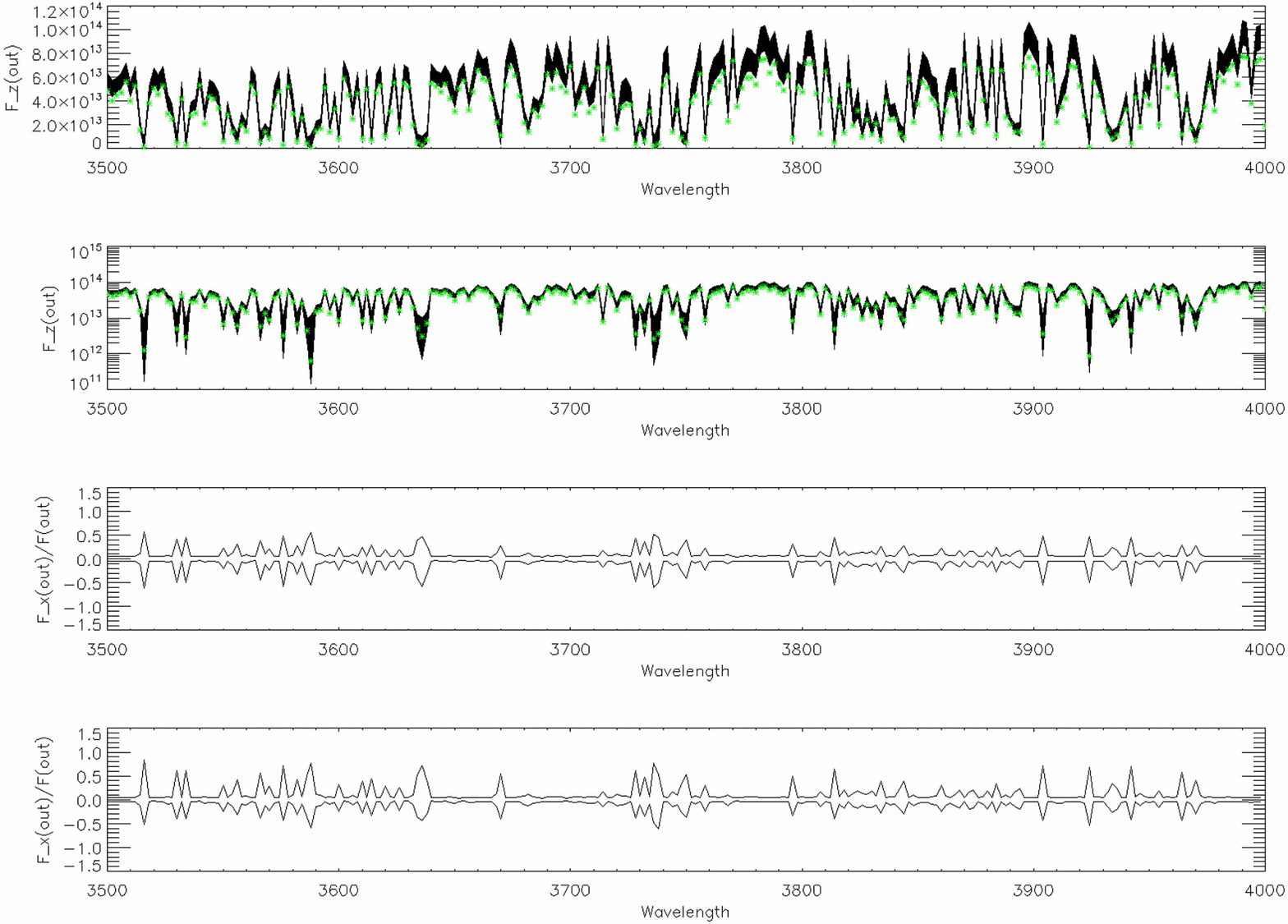}}
\caption{\label{fig:HGL3500}
Comparison between the flux vectors across the outermost voxels for the \phxT\
UV spectra computed for the 3D hydro structure and the \phxO\ spectrum for the G2V
dwarf test model ($\Teff=5700\K$, $\logg=4.5$).  In the \phxT\ calculations we
have used a 3D coordinate system with a total of $141\times 141 \times 151$
Cartesian grid points in $x$, $y$, and $z$, respectively, the periodic boundary
conditions are set in the (horizontal) $x,y$ plane. The 3D radiative transport
equation is solved for $n_{\theta}=64$ and $n_{\phi}=64$ solid angle points.
The top panels show the $F_z$ component of all outer voxels in linear and
logarithmic scales, respectively, compared to the results of the 1D comparison
model. 
{The bottom panels show the corresponding maxima and minima  of $F_x/|\vec F|$ and
$F_y/|\vec F|$, respectively, over all surface voxels for each wavelength.
Note the difference between this result and that shown in the bottom two panels
of Fig.~\ref{fig:HGL1000}. See text for details.
}
The wavelengths are given in {\AA} and the fluxes are in cgs units.
}
\end{figure*}

\begin{figure*}
\centering
\resizebox{\hsize}{!}{\includegraphics[angle=00]{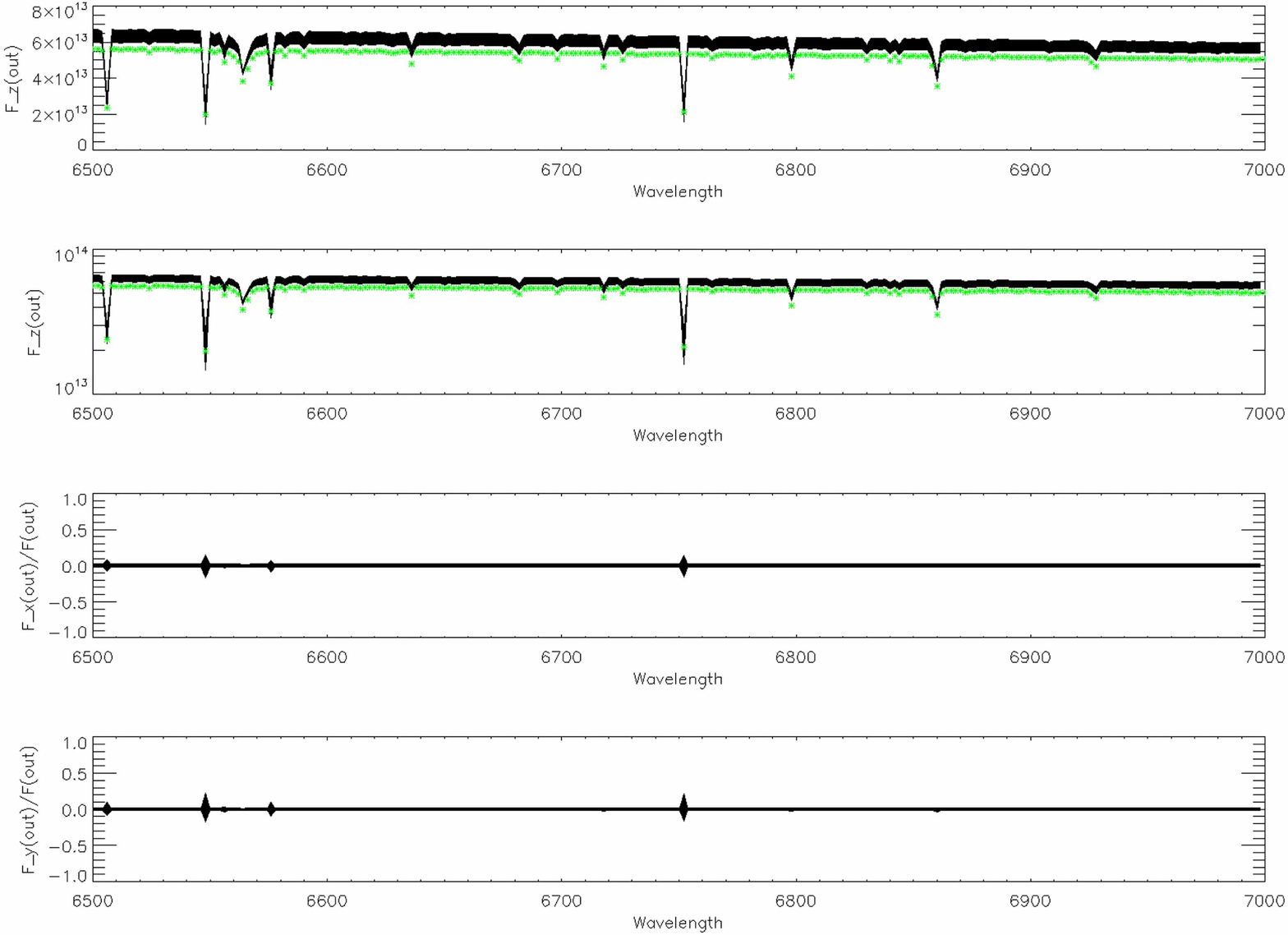}}
\caption{\label{fig:HGL6500}
Comparison between the flux vectors across the outermost voxels for the \phxT\
optical spectra computed for the 3D hydro structure and the \phxO\ spectrum for the G2V
dwarf test model ($\Teff=5700\K$, $\logg=4.5$).  In the \phxT\ calculations we
have used a 3D coordinate system with a total of $141\times 141 \times 151$
Cartesian grid points in $x$, $y$, and $z$, respectively, the periodic boundary
conditions are set in the (horizontal) $x,y$ plane. The 3D radiative transport
equation is solved for $n_{\theta}=64$ and $n_{\phi}=64$ solid angle points.
The top panels show the $F_z$ component of all outer voxels in linear and
logarithmic scales, respectively, compared to the results of the 1D comparison
model.  The bottom panels show the corresponding runs of $F_x/|\vec F|$ and
$F_y/|\vec F|$, respectively. {See text for details.}
The wavelengths are given in {\AA} and the fluxes are in cgs units.
}
\end{figure*}

We show example results in Figs. \ref{fig:HGL1000} to \ref{fig:HGL6500} in terms of the
$x$, $y$, and $z$ components of the flux vectors of each outer boundary voxel.
The $F_z$ components are, in addition, compared to the 1D model for the G2V star
with the parameters $\Teff=5700\K,\ \logg=4.5$ ($*$
symbols in the figures). The general shape of the 3D spectra compare well to
the 1D solar type model, of course there are large  variations across the horizontal
plane. In the UV the differences are largest, a number of voxel flux vectors show
strong line  emission, whereas the radiative+convective equilibrium  1D model only
shows absorption features. This is to be expected as the 3D simulation of
convection gives significant temperature variations across the volume
considered, in particular in the horizontal plane. These variations have
considerable effect on the radiative transfer solution: The horizontal
components of the flux vectors of each voxel compared to the length of the flux
vector $F=|\vec F|$, $F_x/F$ and $F_y/F$, show quite substantial variations for
different wavelengths. The variations are much larger for smaller wavelengths
(e.g., in the UV), due to the larger temperature dependence of the source
functions for smaller wavelengths which translates to {larger} horizontal flux
components  for small wavelengths compared to longer wavelengths. 

\begin{figure*}
\centering
\resizebox{\hsize}{!}{\includegraphics[angle=180]{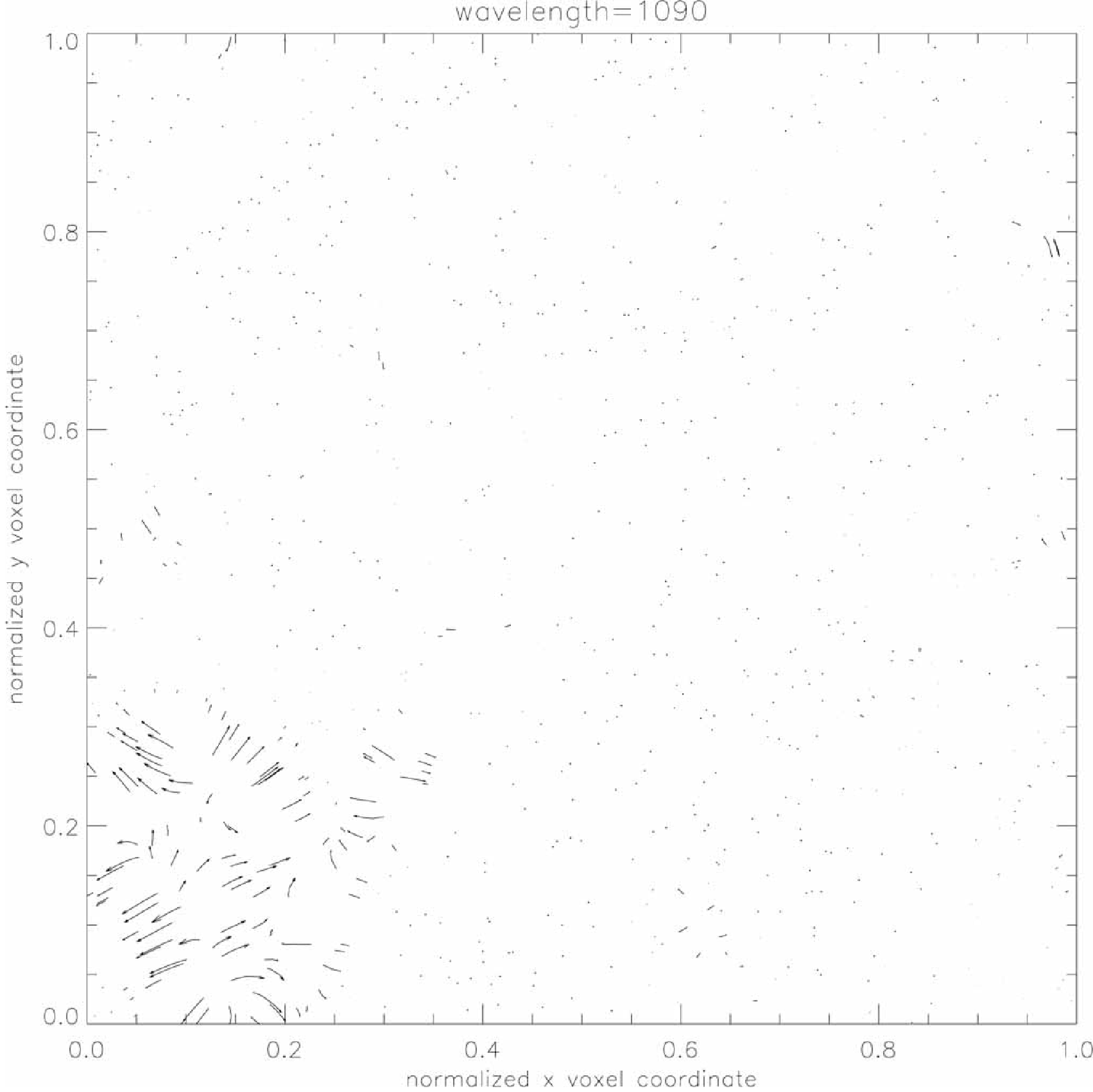}}
\caption{\label{fig:HGL1090_flow}
Illustration of horizontal energy flow for the outermost voxels of the 
the 3D hydro structure for the red spectral range.
{The graphics shows
the flowlines of the $x$ and $y$ components of the flux vector $\vec{F}$.
Here, a flowline connects points of constant $|(F_x,F_y)|$ following 
the direction of $(F_x,F_y)$.}
The 3D radiative transport
equation is solved for $n_{\theta}=64$ and $n_{\phi}=64$ solid angle points.
The wavelengths are given in {\AA}. {The normalized $x$ and
$y$ voxel coordinates are shown on the $x$ and $y$ axes, respectively.}
}
\end{figure*}

\begin{figure*}
\centering
\resizebox{\hsize}{!}{\includegraphics[angle=180]{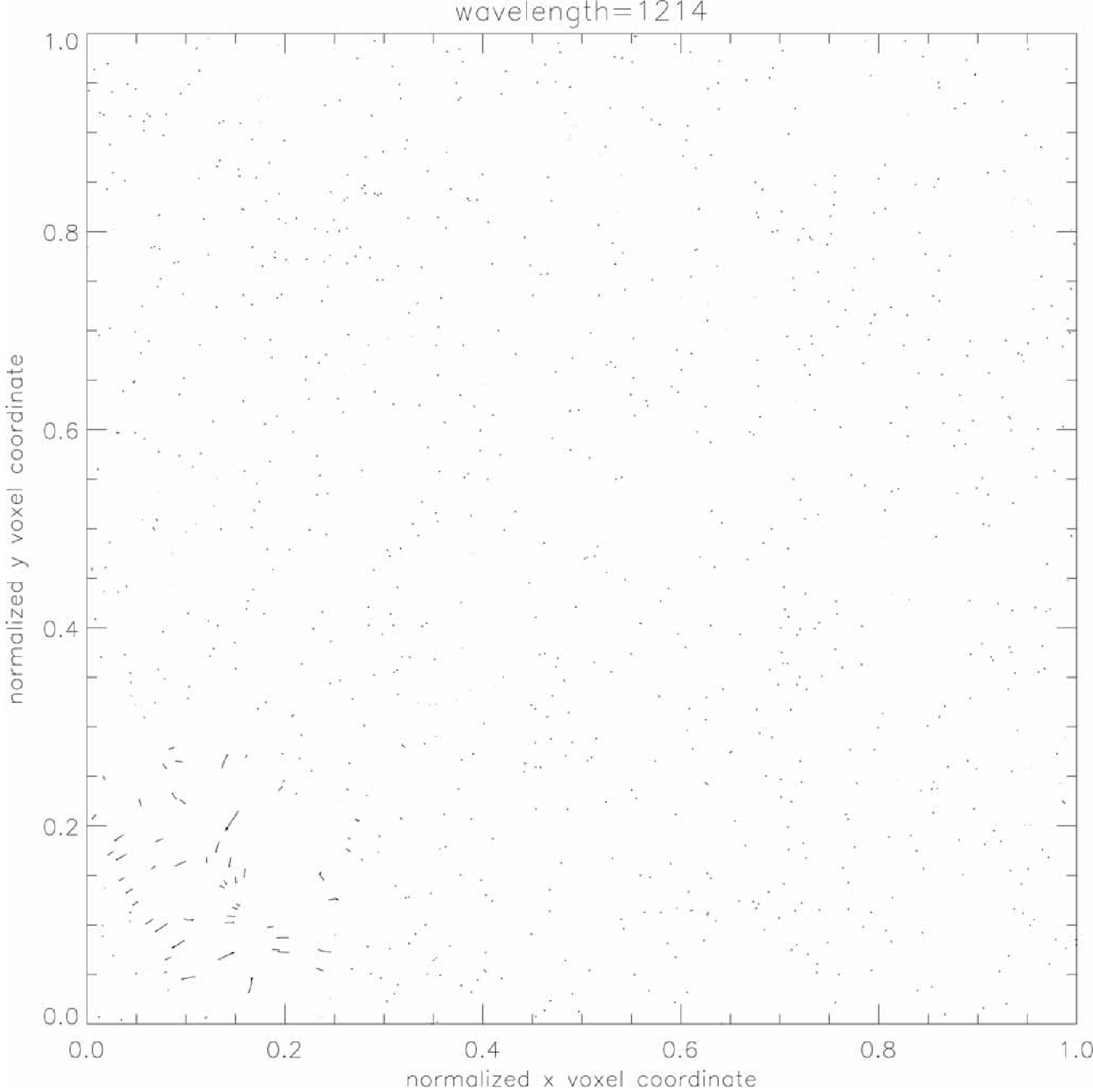}}
\caption{\label{fig:HGL1214_flow}
Illustration of horizontal energy flow for the outermost voxels of the 
the 3D hydro structure for the red spectral range.
{The graphics shows
the flowlines of the $x$ and $y$ components of the flux vector $\vec{F}$.
Here, a flowline connects points of constant $|(F_x,F_y)|$ following 
the direction of $(F_x,F_y)$.}
The 3D radiative transport
equation is solved for $n_{\theta}=64$ and $n_{\phi}=64$ solid angle points.
The wavelengths are given in {\AA}. {The normalized $x$ and
$y$ voxel coordinates are shown on the $x$ and $y$ axes, respectively.}
}
\end{figure*}

\begin{figure*}
\centering
\resizebox{\hsize}{!}{\includegraphics[angle=180]{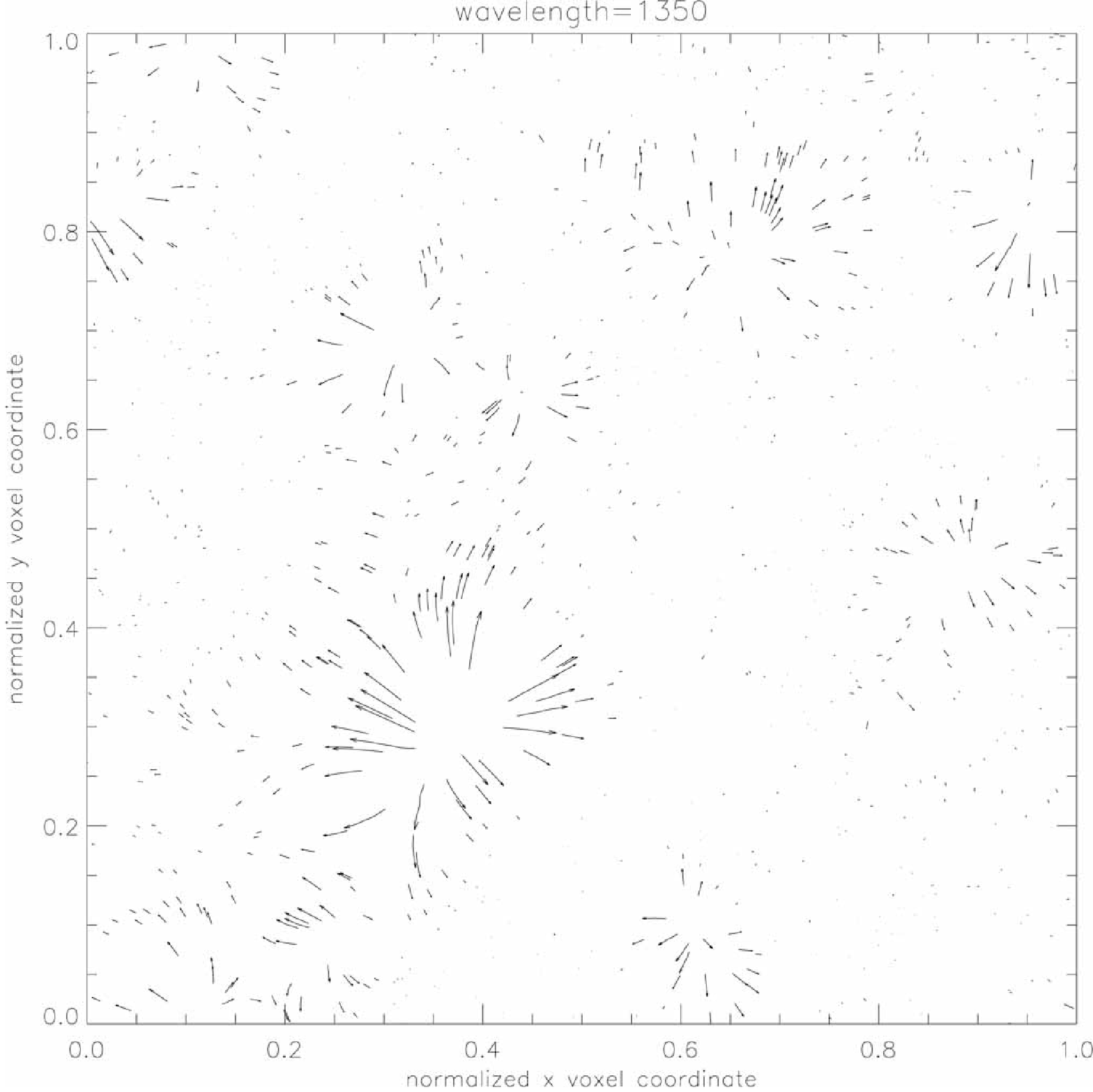}}
\caption{\label{fig:HGL1350_flow}
Illustration of horizontal energy flow for the outermost voxels of the 
the 3D hydro structure for the red spectral range.
{The graphics shows
the flowlines of the $x$ and $y$ components of the flux vector $\vec{F}$.
Here, a flowline connects points of constant $|(F_x,F_y)|$ following 
the direction of $(F_x,F_y)$.}
The 3D radiative transport
equation is solved for $n_{\theta}=64$ and $n_{\phi}=64$ solid angle points.
The wavelengths are given in {\AA}. {The normalized $x$ and
$y$ voxel coordinates are shown on the $x$ and $y$ axes, respectively.}
}
\end{figure*}

\begin{figure*}
\centering
\resizebox{\hsize}{!}{\includegraphics[angle=180]{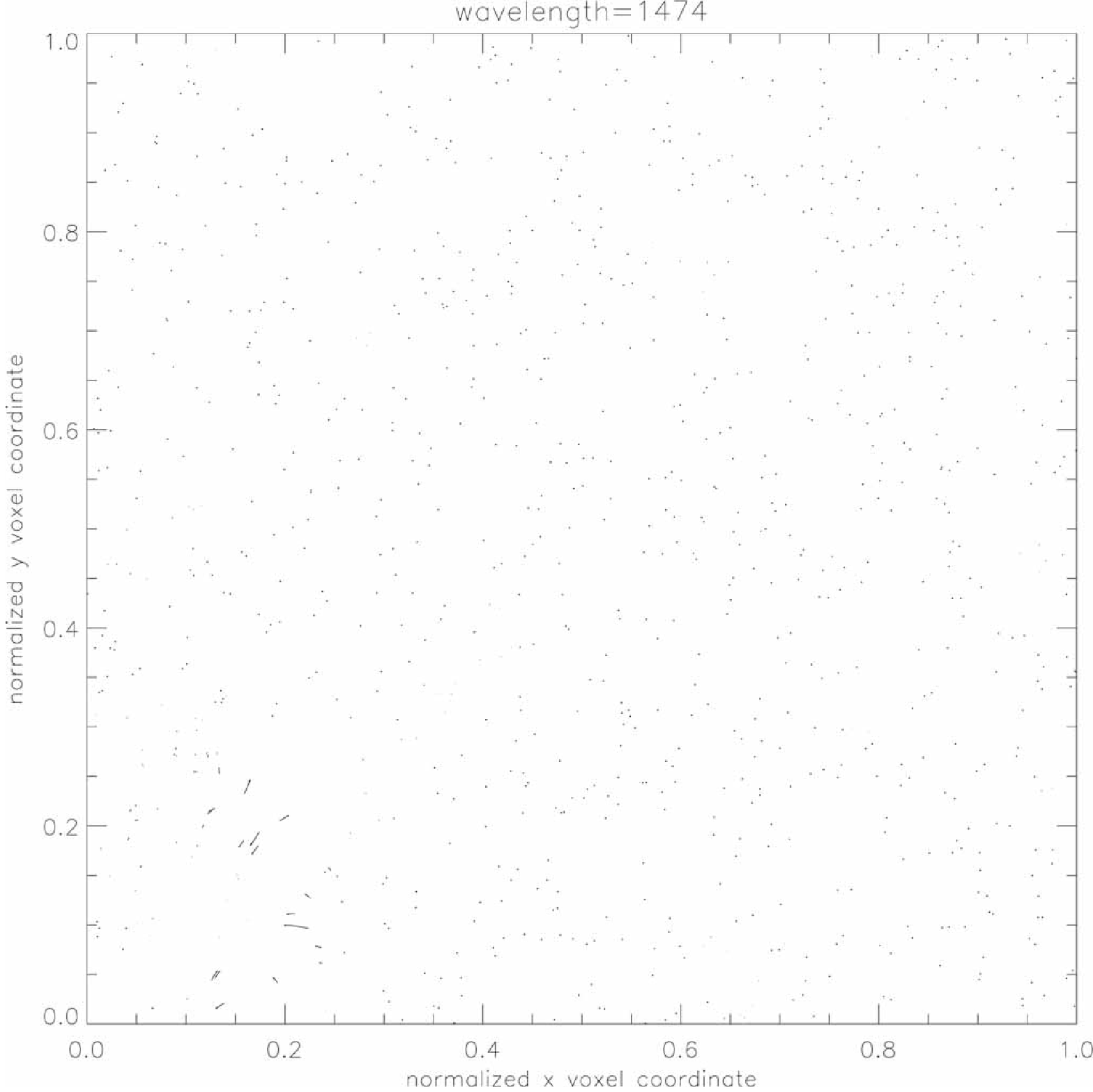}}
\caption{\label{fig:HGL1474_flow}
Illustration of horizontal energy flow for the outermost voxels of the 
the 3D hydro structure for the red spectral range.
{The graphics shows
the flowlines of the $x$ and $y$ components of the flux vector $\vec{F}$.
Here, a flowline connects points of constant $|(F_x,F_y)|$ following 
the direction of $(F_x,F_y)$.}
The 3D radiative transport
equation is solved for $n_{\theta}=64$ and $n_{\phi}=64$ solid angle points.
The wavelengths are given in {\AA}. {The normalized $x$ and
$y$ voxel coordinates are shown on the $x$ and $y$ axes, respectively.}
}
\end{figure*}

\begin{figure*}
\centering
\resizebox{\hsize}{!}{\includegraphics[angle=180]{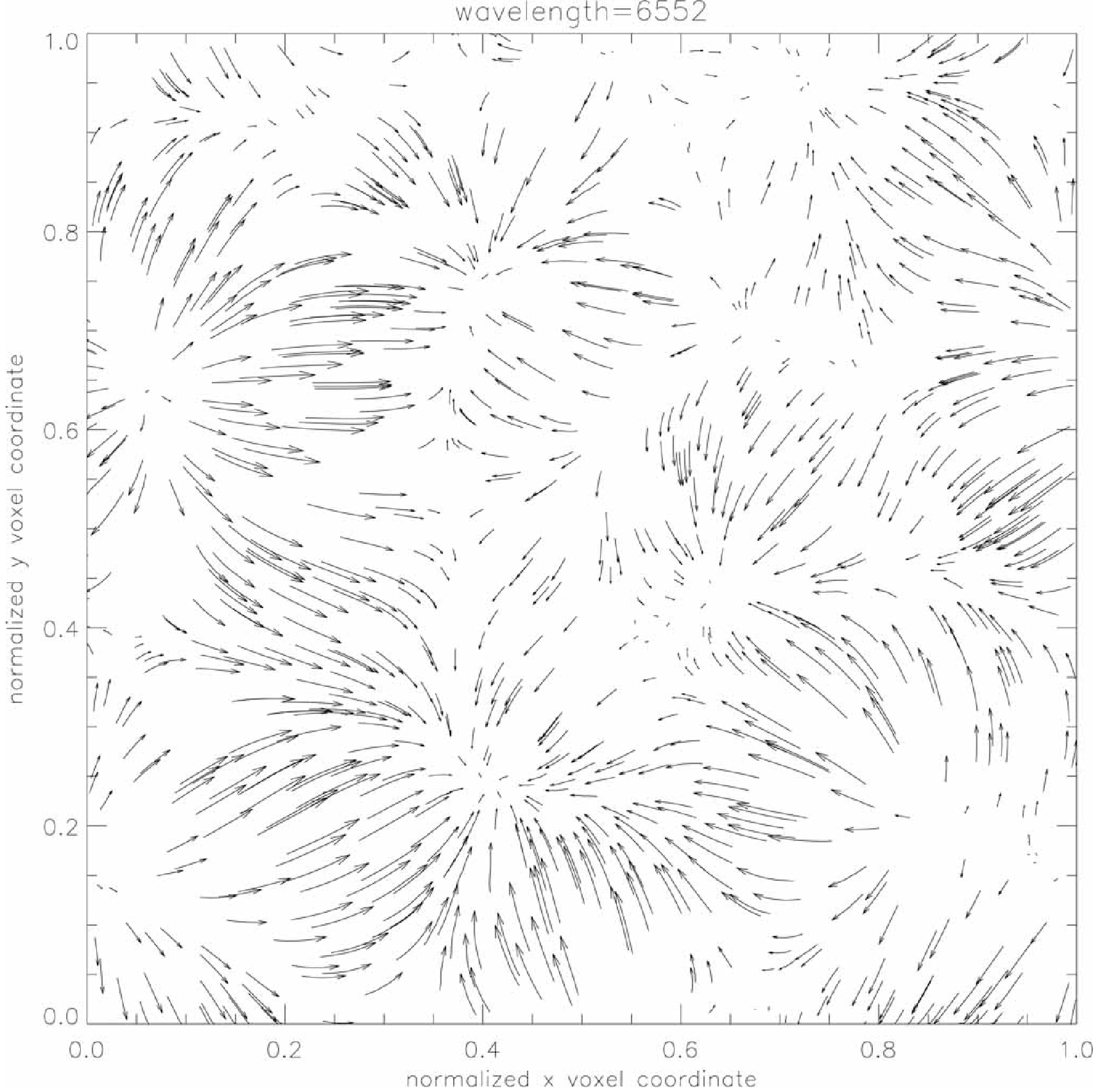}}
\caption{\label{fig:HGL6552_flow}
Illustration of horizontal energy flow for the outermost voxels of the 
the 3D hydro structure for the red spectral range.
{The graphics shows
the flowlines of the $x$ and $y$ components of the flux vector $\vec{F}$.
Here, a flowline connects points of constant $|(F_x,F_y)|$ following 
the direction of $(F_x,F_y)$.}
The 3D radiative transport
equation is solved for $n_{\theta}=64$ and $n_{\phi}=64$ solid angle points.
The wavelengths are given in {\AA}. {The normalized $x$ and
$y$ voxel coordinates are shown on the $x$ and $y$ axes, respectively.}
}
\end{figure*}

\begin{figure*}
\centering
\resizebox{\hsize}{!}{\includegraphics[angle=180]{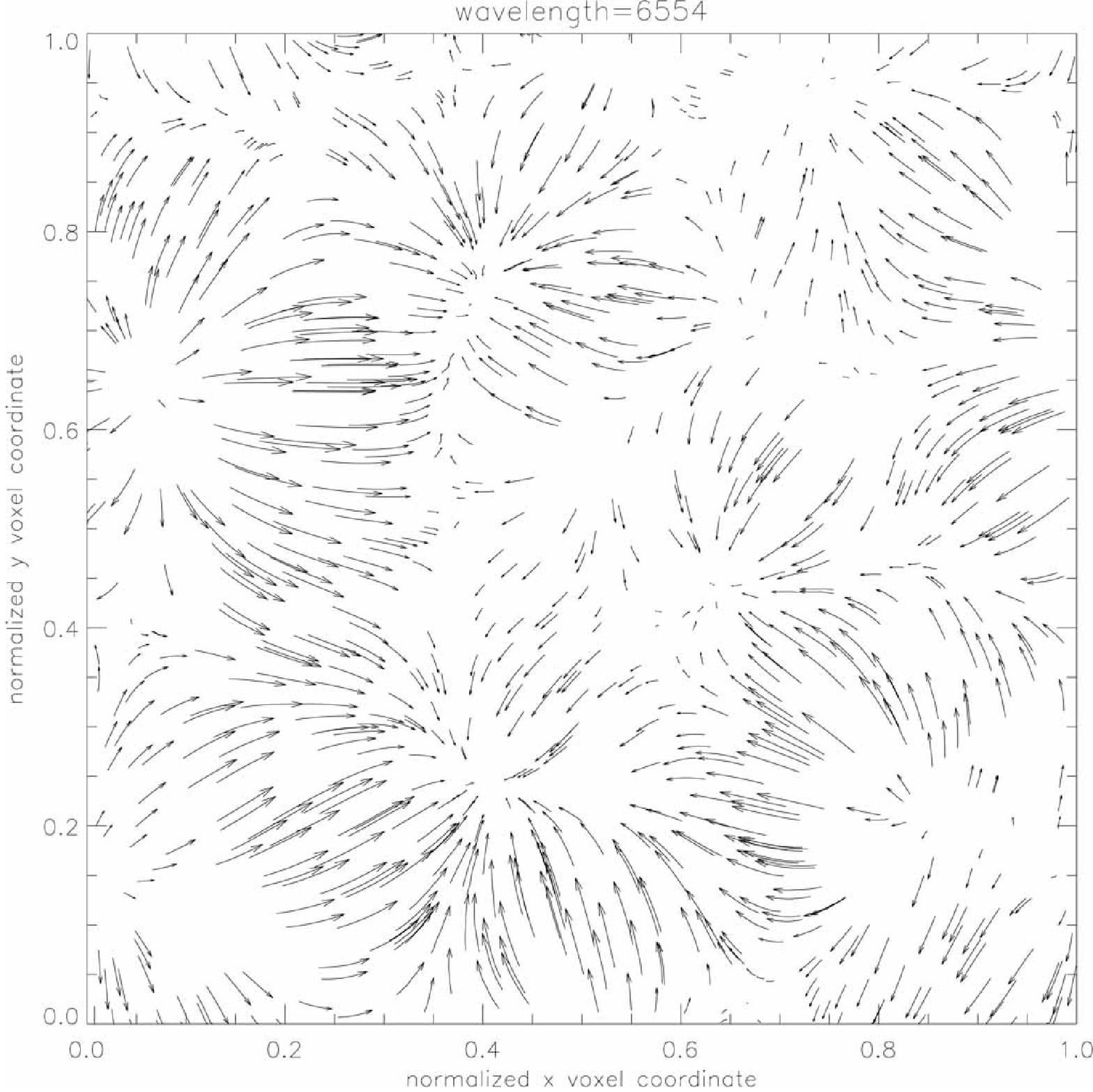}}
\caption{\label{fig:HGL6554_flow}
Illustration of horizontal energy flow for the outermost voxels of the 
the 3D hydro structure for the red spectral range.
{The graphics shows
the flowlines of the $x$ and $y$ components of the flux vector $\vec{F}$.
Here, a flowline connects points of constant $|(F_x,F_y)|$ following 
the direction of $(F_x,F_y)$.}
The 3D radiative transport
equation is solved for $n_{\theta}=64$ and $n_{\phi}=64$ solid angle points.
The wavelengths are given in {\AA}. {The normalized $x$ and
$y$ voxel coordinates are shown on the $x$ and $y$ axes, respectively.}
}
\end{figure*}

\begin{figure*}
\centering
\resizebox{\hsize}{!}{\includegraphics[angle=180]{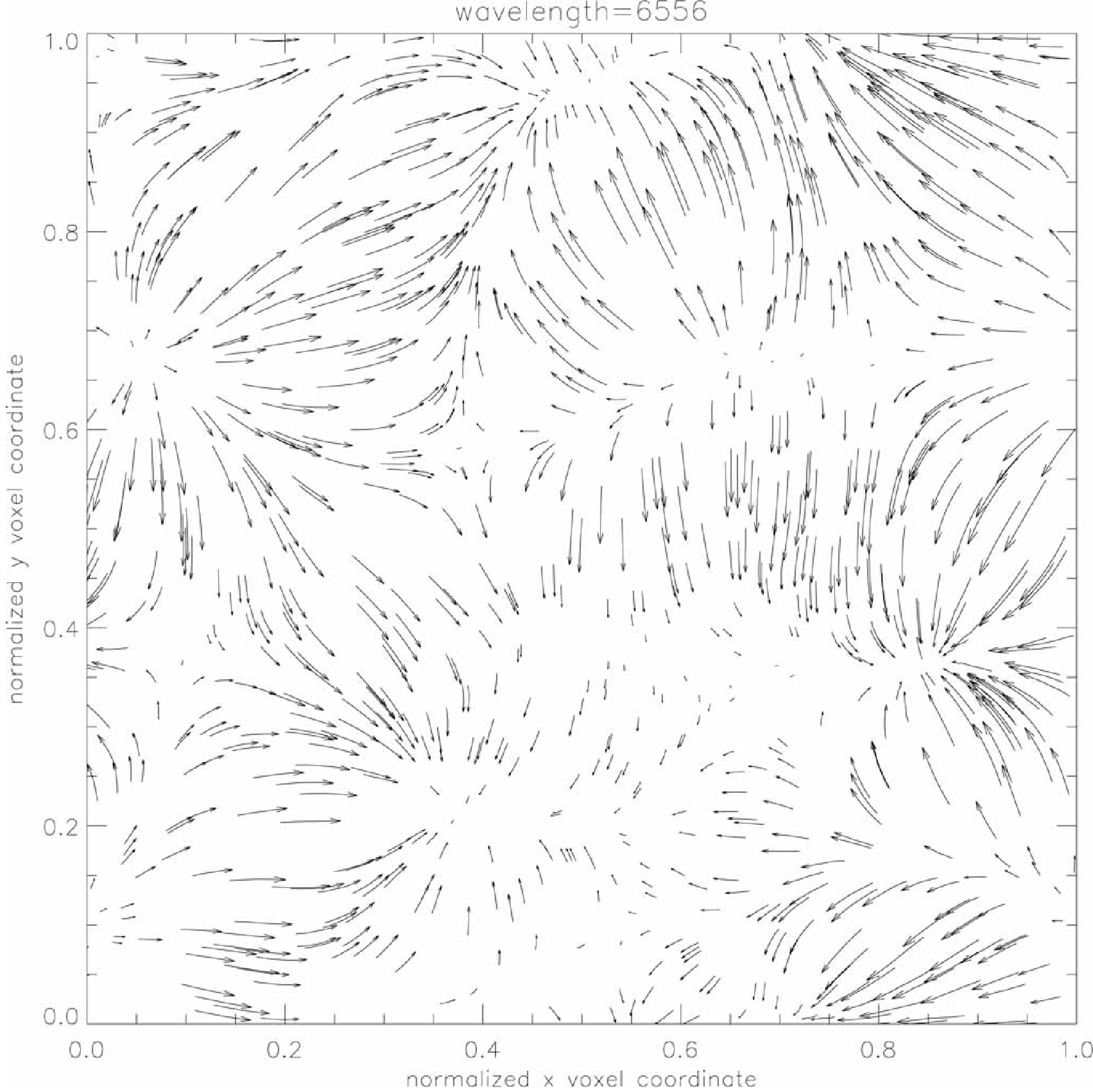}}
\caption{\label{fig:HGL6556_flow}
Illustration of horizontal energy flow for the outermost voxels of the 
the 3D hydro structure for the red spectral range.
{The graphics shows
the flowlines of the $x$ and $y$ components of the flux vector $\vec{F}$.
Here, a flowline connects points of constant $|(F_x,F_y)|$ following 
the direction of $(F_x,F_y)$.}
The 3D radiative transport
equation is solved for $n_{\theta}=64$ and $n_{\phi}=64$ solid angle points.
The wavelengths are given in {\AA}. {The normalized $x$ and
$y$ voxel coordinates are shown on the $x$ and $y$ axes, respectively.}
}
\end{figure*}

\begin{figure*}
\centering
\resizebox{\hsize}{!}{\includegraphics[angle=180]{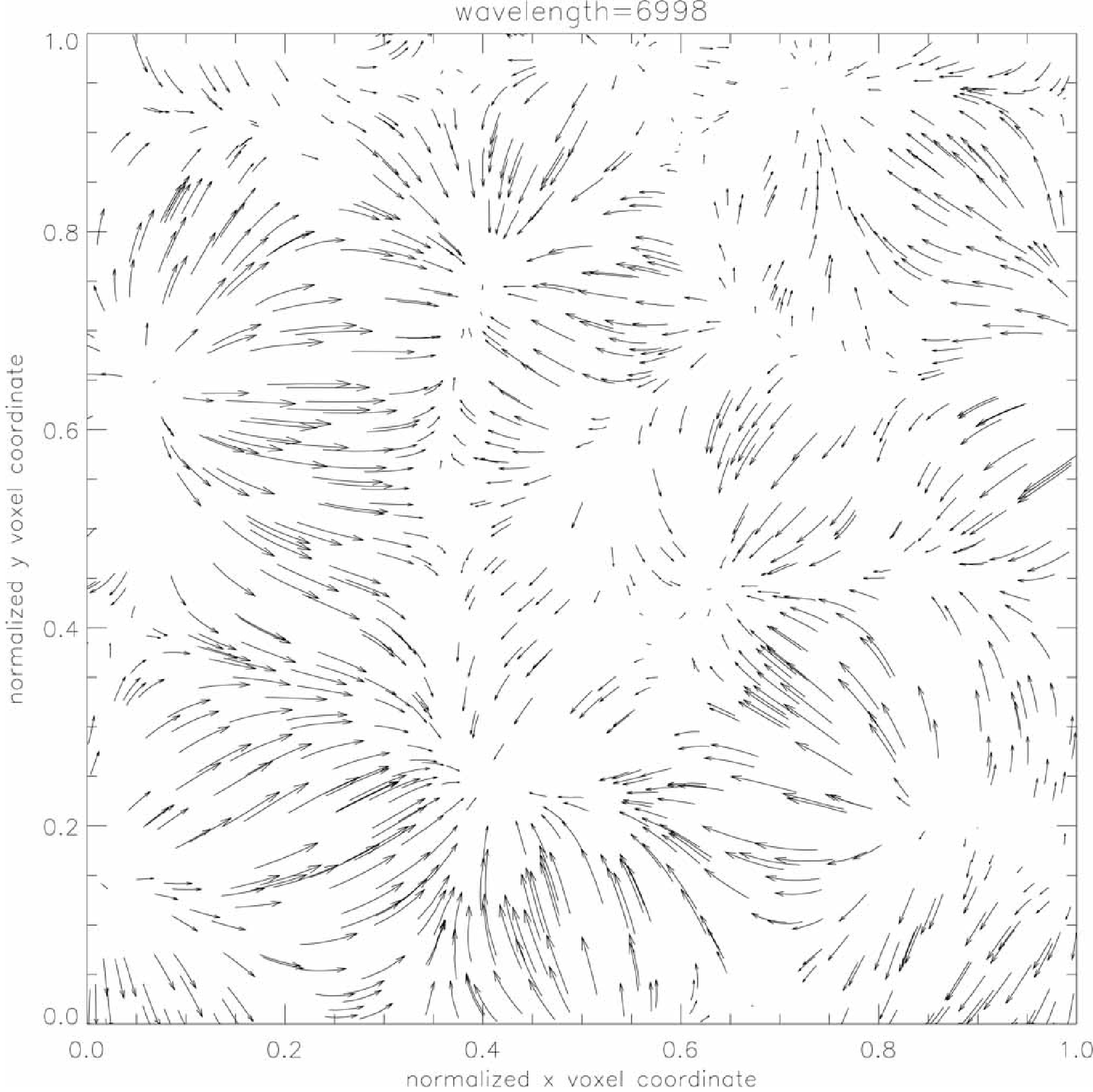}}
\caption{\label{fig:HGL6998_flow}
Illustration of horizontal energy flow for the outermost voxels of the 
the 3D hydro structure for the red spectral range.
{The graphics shows
the flowlines of the $x$ and $y$ components of the flux vector $\vec{F}$.
Here, a flowline connects points of constant $|(F_x,F_y)|$ following 
the direction of $(F_x,F_y)$.}
The 3D radiative transport
equation is solved for $n_{\theta}=64$ and $n_{\phi}=64$ solid angle points.
The wavelengths are given in {\AA}. {The normalized $x$ and
$y$ voxel coordinates are shown on the $x$ and $y$ axes, respectively.}
}
\end{figure*}

{
The components of the flux vectors in the $x-y$ plane can be larger than the
$z$ component, strongly dependent on the wavelength and on the location of the
voxel. This is illustrated in Figs.~\ref{fig:HGL1090_flow} --
\ref{fig:HGL6998_flow}, which show the flowlines of the $x-y$ components of
the flux vector at the surface. The flow distances are much larger at optical wavelengths
than in the UV due to the larger UV opacities. 
The `pattern' of the horizontal energy flow depends strongly on the wavelength,
it is also significantly different in the cores of strong lines compared to the
surrounding continuum. The horizontal heat exchange could have in turn
noticeable effects on the gas flow pattern. 
}

\subsection{Supernovae}

\begin{figure*}
\centering
\resizebox{\hsize}{!}{\includegraphics[angle=00]{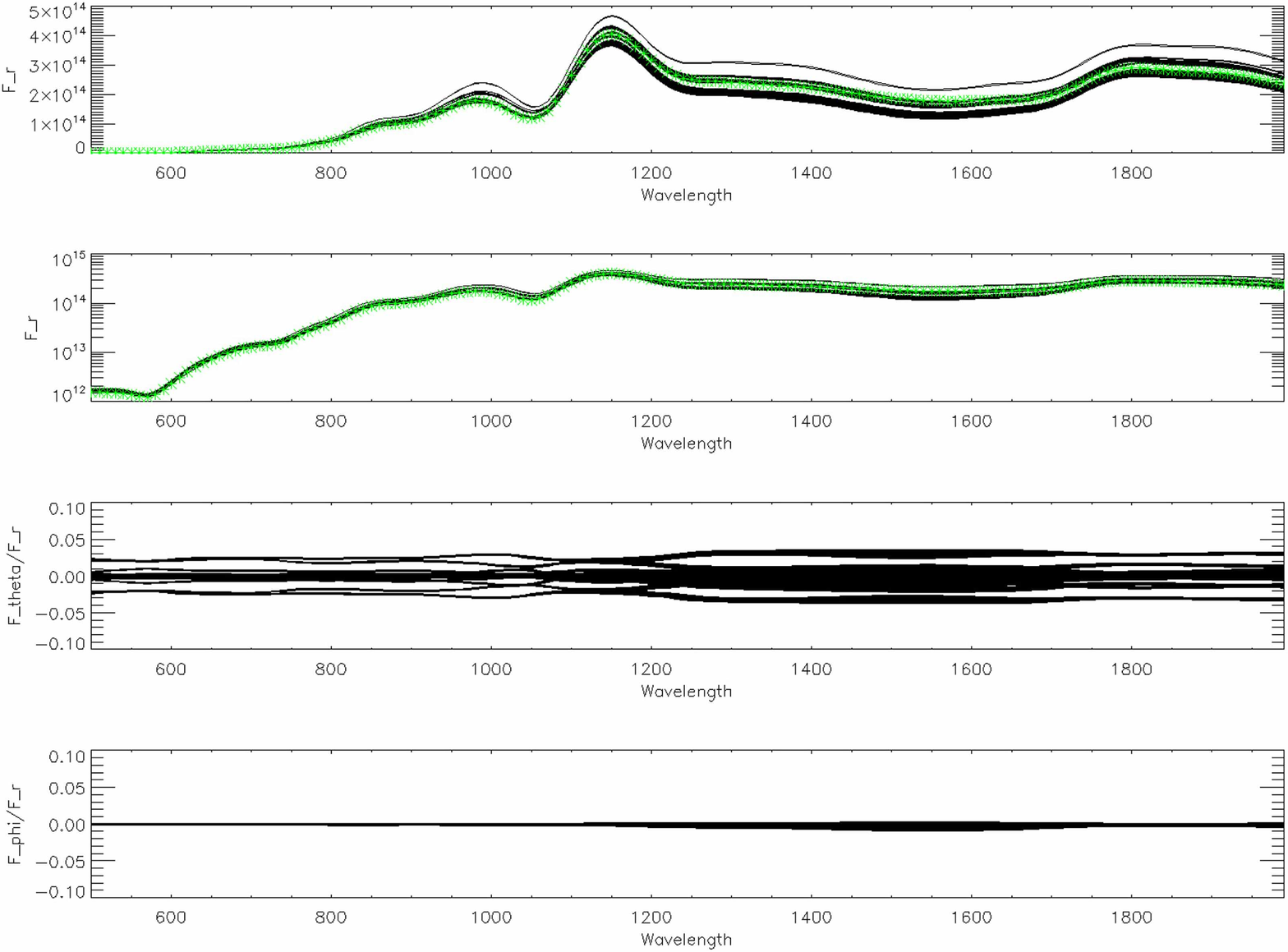}}
\caption{\label{fig:SN:CMF} 
Comparison between the \phxO\ co-moving frame UV spectrum (computed with 256
layers, $*$ symbols) and the co-moving frame flux vectors across the outermost
voxels for the \phxT\ spectra computed for the Supernova test model.  In the
\phxT\ calculations we have used a 3D spherical coordinate system with $n_r =
129$, $n_{\theta_c}=33$ and $n_{\phi_c}=65$ points for a total of about 275k
voxels. The calculations used $128^2$ solid angle points.  The top panels show
the $F_r$ component of all outer voxels in linear and logarithmic scales,
respectively. The bottom panels show the corresponding runs of $F_\theta/F_r$
and $F_\phi/F_r$, respectively.  The should be identically zero and the
deviations measure the internal accuracy.  The wavelengths are given in {\AA}
and the fluxes are in cgs units.
}
\end{figure*}

The modeling of supernova spectra is a very important
application of \phxT\ modeling as it is expected, and explosion
models show, that supernova explosions are intrinsically 3D driven.
For the calculations shown here we use the Lagrangian frame 3DRT in
spherical 3D coordinates as 
discussed in paper V. The test model is a simplified model for 
a type II supernova atmosphere with a maximum expansion speed of about 
$0.13c$. The model is a simple uniform composition model with the
density parameterized as $\rho \propto r^{-9}$, and a ``photospheric
velocity'' of $v_0 = 7600$~km~s$^{-1}$, and a model temperature of
$T_{\mathrm{model}} = 17000$~K. These conditions correspond roughly to
those of SN~1999em seven days after explosion.
In Fig.~\ref{fig:SN:CMF} we show the CMF spectrum of the \phxT\ run
compared to the corresponding \phxO\ synthetic spectrum. Due to computer
time limitations we could only run a relatively small 3D model
with $n_r=129$,  $n_{\theta_c}=33$ and $n_{\phi_c}=65$ points
and $128^2$ solid angle points. The small angular resolution
causes the scatter in the $F_r$ plots and the errors in the 
$F_{\theta_c}$ and $F_{\phi_c}$ components. In general the 
agreement is acceptable for this test run, for a full scale
3D SN spectrum the resolution in $(\theta_c,\phi_c)$ should
be increased to $(65,129)$ at least and the angular resolution
should be at least $512^2$ (which reduces the bandwidth dramatically,
see Paper V).

\section{Summary and Conclusions}

We have described first results we have obtained by incorporating
the 3D radiative transfer framework we have discussed in Papers I-V
into our general purpose model atmosphere package \phx, thus allowing
both 1D models (\phxO) and 3D models (\phxT) with the same micro-physics. 
We have verified and tested \phxT\ by computing a number of test
spectra for 1D conditions and comparing the results to the 
corresponding \phxO\ calculations. The conditions range from M dwarfs,
solar type stars to A stars and Type II supernovae with relativistic
expansions speeds. In addition, we have calculated spectra for a 3D 
hydrodynamical simulation of solar atmosphere convection.
These tests demonstrate the it is now possible to calculate
realistic spectra for 3D configurations including complex micro-physics.
\phxT\ can be used to calculate synthetic spectra for a number of 
complex 3D atmosphere model, including irradiated stars or planets, 
novae, and supernovae. We are currently working on extensions
of the 3D radiative transfer framework to arbitrary velocity fields
in the Euler (for low velocities, e.g., in convection simulations or
planetary winds) and the Lagrangian (for Supernovae, accretion disks and
matter flow in the vicinity of black holes) frames, which will extend
the applications of \phxT\ significantly.

\begin{acknowledgements}
This work was supported in part by 
DFG GrK 1351 and SFB 676, as well as 
NSF grant AST-0707704, and US DOE Grant DE-FG02-07ER41517.
The calculations presented here were performed at the H\"ochstleistungs
Rechenzentrum Nord (HLRN); at the Hamburger Sternwarte IBM Regatta Systems,
Apple G5, and Delta Opteron clusters financially supported by the DFG and the
State of Hamburg; and at the National Energy Research Supercomputer Center
(NERSC), which is supported by the Office of Science of the U.S.  Department of
Energy under Contract No. DE-AC03-76SF00098.  We thank all these institutions
for a generous allocation of computer time.
\end{acknowledgements}

\bibliography{yeti,radtran,rte_paper2,stars}

\clearpage

\begin{table*}
\caption{\label{tab:1} Strong scaling behavior of a M dwarf model test case for
different configurations and total number of processors used. See text for details.}

\begin{centering}
\begin{tabular}{rrrrrrrrrrr}
           &            &            &    Timing: &            &            &            &            &   Scaling: &            &            \\
    n(MPI) & cluster size & n(cluster) & line opacity &       3DRT &       Comm &      total &            & Line opacity &       3DRT &      total \\
      2048 &        512 &          4 &       2473 &       9942 &        436 &      12909 &            &     96.7\% &     79.3\% &     81.9\% \\
      2048 &        256 &          8 &       2414 &       8054 &        285 &      10789 &            &     99.1\% &     97.9\% &     98.0\% \\
      1024 &        512 &          2 &       4900 &      20103 &        754 &      25872 &            &     97.6\% &     78.4\% &     81.7\% \\
      1024 &        256 &          4 &       4741 &      15966 &        604 &      21385 &            &    100.9\% &     98.7\% &     98.9\% \\
      1024 &        128 &          8 &       4681 &      14275 &        471 &      19483 &            &    102.2\% &    110.4\% &    108.6\% \\
       512 &        512 &          1 &       9780 &      39793 &       1539 &      51341 &            &     97.8\% &     79.2\% &     82.4\% \\
       512 &        256 &          2 &       9566 &      31524 &       1061 &      42300 &            &    100.0\% &    100.0\% &    100.0\% \\
\end{tabular}  
\end{centering}
\end{table*}

\clearpage

\end{document}